\documentclass{pspum-l}
\usepackage{amssymb, amscd}
\begin{document}

\font\nt=cmr7
\def\note#1
{\marginpar
{\nt $\leftarrow$
\par
\hfuzz=20pt \hbadness=9000 \hyphenpenalty=-100 \exhyphenpenalty=-100
\pretolerance=-1 \tolerance=9999 \doublehyphendemerits=-100000
\finalhyphendemerits=-100000 \baselineskip=6pt
#1}\hfuzz=1pt}

\def\note#1{}

\newcommand{\bignote}[1]{\begin{quote} \sf #1 \end{quote}}

\def\I {\mathbb{1}}
\def\Z {\mathbb{Z}}
\def\pRe{\mathbbb{R}}
\def\Re {\mathbb{R}}
\def\C {\mathbb{C}}
\def\pC{\mathbbb{C}}

\def\H {\mathbb{H}}
\def\ni{\noindent}
\newcommand\noi{\noindent}
\def\nn{\nonumber}
\newcommand\seq{\;\;=\;\;}

\def\eq{\begin{equation}}
\def\eqe{\end{equation}}
\def\eqa{\begin{eqnarray}}
\def\eqae{\end{eqnarray}}
\def\bea{\begin{eqnarray}}
\def\ena{\end{eqnarray}}

\def\st{\star}
\def\dZ2p{\frac{dZ_1}{2\p i}}

\newcommand{\be}{\begin{equation}}
\newcommand{\ee}{\end{equation}}
\newcommand{\ba}{\begin{eqnarray}}
\newcommand{\ea}{\end{eqnarray}}
\newcommand{\al}{\mbox{$\alpha$}}
\newcommand{\als}{\mbox{$\alpha_{s}$}}
\newcommand{\s}{\mbox{$\sigma$}}
\newcommand{\lm}{\mbox{$\mbox{ln}(1/\alpha)$}}
\newcommand{\bi}[1]{\bibitem{#1}}
\newcommand{\fr}[2]{\frac{#1}{#2}}
\newcommand{\sv}{\mbox{$\vec{\sigma}$}}
\newcommand{\gm}{\mbox{$\gamma_{\mu}$}}
\newcommand{\gn}{\mbox{$\gamma_{\nu}$}}
\newcommand{\Le}{\mbox{$\fr{1+\gamma_5}{2}$}}
\newcommand{\R}{\mbox{$\fr{1-\gamma_5}{2}$}}
\newcommand{\GD}{\mbox{$\tilde{G}$}}
\newcommand{\gf}{\mbox{$\gamma_{5}$}}
\newcommand{\om}{\mbox{$\omega$}}
\newcommand{\Ima}{\mbox{Im}}\newtheorem{thm}{Theorem}[subsection]
\newtheorem{mthm}{Theorem}

\newcommand{\Rea}{\mbox{Re}}
\newcommand{\Tr}{\mbox{Tr}}
\newtheorem{cor}[thm]{Corollary}
\newtheorem{prop}[thm]{Proposition}
\newtheorem{lemma}[thm]{Lemma}
\newtheorem{claim}[thm]{Claim}
\newtheorem{conj}[thm]{Conjecture}
\newtheorem{definition}[thm]{Definition}
\newtheorem{proposition}[thm]{Proposition}
\newtheorem{condition}[thm]{Condition}
\numberwithin{equation}{section}

\def\a{\alpha}
\def\b{\beta}
\def\c{\chi}
\def\d{\delta}
\def\e{\epsilon}           % Also, \varepsilon
\def\f{\phi}               %      \varphi
\def\g{\gamma}
\def\h{\eta}
\def\i{\iota}
\def\j{\psi}
\def\k{\kappa}                    % Also, \varkappa (see below)
\def\l{\lambda}
\def\m{\mu}
\def\n{\nu}
\def\o{\omega}
\def\p{\pi}                % Also, \varpi
\def\q{\theta}                    %     \vartheta
\def\r{\rho}                      %     \varrho
\def\s{\sigma}                    %     \varsigma
\def\t{\tau}
\def\u{\upsilon}
\def\x{\xi}
\def\z{\zeta}
\def\D{\Delta}
\def\F{\Phi}
\def\G{\Gamma}
\def\J{\Psi}
\def\L{\Lambda}
\def\O{\Omega}
\def\P{\Pi}
\def\Q{\Theta}
\def\S{\Sigma}
\def\U{\Upsilon}
\def\X{\Xi}
\def\del{\partial}
\def\pa{\partial}

% Calligraphic letters

\def\ca{{\mathcal A}}
\def\cb{{\mathcal B}}
\def\cc{{\mathcal C}}
\def\cd{{\mathcal D}}
\def\ce{{\mathcal E}}
\def\cf{{\mathcal F}}
\def\cg{{\mathcal G}}
\def\ch{{\mathcal H}}
\def\ci{{\mathcal I}}
\def\cj{{\mathcal J}}
\def\ck{{\mathcal K}}
\def\cl{{\mathcal L}}
\def\cm{{\mathcal M}}
\def\cn{{\mathcal N}}
\def\co{{\mathcal O}}
\def\cp{{\mathcal P}}
\def\cq{{\mathcal Q}}
\def\car{{\mathcal R}}
\def\cs{{\mathcal S}}
\def\ct{{\mathcal T}}
\def\cu{{\mathcal U}}
\def\cv{{\mathcal V}}
\def\cw{{\mathcal W}}
\def\cx{{\mathcal X}}
\def\cy{{\mathcal Y}}
\def\cz{{\mathcal Z}}

\def\vecnab{\vec{\nabla}}
\def\vx{\vec{x}}
\def\vy{\vec{y}}
\def\arrowk{\stackrel{\rightarrow}{k}}
\def\kbar{k\!\!\!^{-}}
\def\karrow{k\!\!\!{\rightarrow}}
\def\arrowl{\stackrel{\rightarrow}{\ell}}
\def\var{\varphi}

%\fig 3.15in by 3.88in (15 scaled 800) make figure smaller
%Use this when scanning figures into an equation file - example -\fig 1.24in by
%1.31in (3.4)
%\def\fig #1 by #2 (#3){\vbox to #2{
 %   \hrule width #1 height 0pt depth 0pt\vfill\special{picture #3}}}

% Usage: #1 = width, #2 = height (see pictures window),
%        #3 = name of picture, or name "scaled" magnification ¥ 1000
%
% For best bitmaps, do at 72dpi, then scale 240

%\def\hook#1{{\vrule height#1pt width0.4pt depth0pt}}
%\def\leftrighthookfill#1{$\mathsurround=0pt \mathord\hook#1
 %       \hrulefill\mathord\hook#1$}
%\def\underhook#1{\vtop{\ialign{##\crcr                 % |_| under
   %     $\hfil\displaystyle{#1}\hfil$\crcr
 %       \noalign{\kern-1pt\nointerlineskip\vskip2pt}
  %      \leftrighthookfill5\crcr}}}

%\def\under#1#2{\mathop{\null#2}\limits_{#1}}

\def\ttZ{\tilde{\tilde{Z}}}
\def\ttz{\tilde{\tilde{z}}}
\def\tz{\tilde{z}}
\def\tZ{\tilde{Z}}
\def\tq{\tilde{\q}}
\def\del{\partial}
\def\half{\frac{1}{2}}
\def\st{\star}
\def\stam{\stackrel{\rightarrow}{\m}}
\def\stab{\stackrel{\rightarrow}{\b}}
\def\half{\frac{1}{2}}
\def\m{\mu}
\def\n{\nu}
\def\g{\gamma}
\def\d{\delta}
\def\in{\infty}
\def\nn{\nonumber\\}
%%%%%%%%%%%%%%%%

\title[Supersymmetry in a simple model]{ Supersymmetry, supergravity, superspace\\ and BRST symmetry in a simple model}
%\title {in a simple model}
\author{ Peter van Nieuwenhuizen}
\address{ C.N. Yang Institute for Theoretical Physics,
Stony Brook University, Stony Brook, NY, 11794-3840, USA}
\copyrightinfo{2000}%            % copyright year
    {American Mathematical Society}% copyright holder
%\dedicatory{To Dennis Sullivan on his 60-th birthday}
\keywords{Supersymmetry, superspace, action, Hamiltonian, spinors, conserved
charges, constraints, gauge fixing, Faddeev-Popov ghosts, supergravity, BRST quantization}
\subjclass[2000]{81T60}
%\date{July, 2002}
\maketitle
%\pagenumbering{roman}
%\mathversion{bold}
\tableofcontents
%\mathversion{normal}

%\pagenumbering{arabic}

%\baselineskip=24pt

%\setcounter{secnumdepth}{-2}
\section{Introduction}

In these lectures we shall introduce rigid supersymmetry,
supergravity (which is the gauge theory of supersymmetry) and
superspace, and apply the results to BRST quantization. We assume
that the reader has never studied these topics. For readers who
want to read more about these ``super" subjects, we give a few
references at the end of this contribution, but the whole point
of these lectures is that one does not need additional references
for a self-contained introduction.  The reader should just sit
down with paper and pencil. We could have decided to begin with
the usual models in $3+1$ dimensional Minkowski space with
coordinates $x$, $y$, $z$ and $t$; this is the standard approach,
but we shall instead consider a much simpler model, with only one
coordinate $t$.  We interpret $t$ as the time coordinate.  For
physicists, the $3+1$ dimensional models are the ones of interest
because they are supposed to describe the real world. For
mathematicians, however, the simpler model may be of more
interest because the basic principles appear without the dressing
of physical complications. Let us begin with three definitions
which should acquire meaning as we go on.

\noindent {\bf Supersymmetry} is a symmetry of the action (to be
explained) with a rigid (constant) anticommuting
parameter%
\footnote{ Technically: a Grassmann variable.  ``Constant'' means 
``independent of the spacetime coordinates $x,y,z$ and $t$''.}  
(usually denoted by $\epsilon$)
between bosonic  (commuting) and fermionic (anticommuting) fields
(again to be explained).  It requires that for every bosonic
particle in Nature there exists a corresponding fermionic
particle, and for every fermionic particle there should exist a
corresponding bosonic particle.  So supersymmetry predicts that
there are twice as many particles as one might have thought.  One
may call these new particles supersymmetric particles.  These
supersymmetric particles will be looked for at CERN (the European
high-energy laboratory) in the coming 8 years.  So far not a
single  %\note{*}
 supersymmetric particle has been discovered: supersymmetry
is a theoretical possibility, but whether Nature is aware of this
possibility remains to be seen.%
\footnote{ This is not the first time a doubling of the number of particles has been
predicted.  In 1931  Dirac predicted that for every fermionic
particle a fermionic antiparticle should exist, and these
antiparticles were discovered in 1932.  We consider in these
lectures only real fields, and  the particles corresponding to
real fields are their own antiparticles.  Hence in these lectures
the notion of antiparticles plays no role.}

\noindent {\bf Supergravity} is the gauge theory of supersymmetry.
 Its action is invariant under transformation rules which depend on a local (space- and time-dependent)
 anticommuting parameter $\epsilon(x,y,z,t)$, and
there is a gauge field for supersymmetry which is called the gravitino field.
 It describes a new hypothetical particle, the gravitino.  The gravitino is the fermionic partner
 of the graviton.  The graviton is the quantum of the gravitional field (also called the
metric).   The astonishing discovery of 1976 was that a gauge
theory of supersymmetry requires gravity: Einstein's 1916 theory
of gravity (called general relativity) is a product of local
supersymmetry.  Phrased differently: local supersymmetry is the
``square root of general relativity'', see~(\ref{moya}). (Likewise,
supersymmetry is the square root of translation symmetry, see
(\ref{Jacobi})).

\noindent {\bf Superspace.} In Nature fields can be divided into
bosonic (commuting) fields and fermionic (anti-commuting)
fields.   This is one of the fundamental discoveries of the
quantum theory of the 1920's.  The anticommuting fields are
described by spinors and the bosonic fields by tensors according
to the spin-statistics theorem of the 1930's.  (Spinors and
tensors refers to their transformation properties under Lorentz
transformations).  One can also introduce in addition to the
usual coordinates $x^\mu$ anticommuting counterparts $\theta^\a$.
The space with coordinates $x$ and $\theta$ is called
superspace.  In the case of a four-dimensional Minkowski space
(our world) there are four coordinates ($x,y,z$ and $t$) and also
four $\theta$'s, namely $\theta^1$, $\theta^2$, $\theta^3$ and
$\theta^4$, but in other dimensions the number of $x$'s and
$\theta$'s are not the same.%
\footnote{ The $x^\mu$ transform as vectors under the Lorentz group while the
$\theta^\a$ transform as spin $1/2$ spinors.}  
 These $\theta$'s
are Grassmann variables \cite{Berezinfa}, for example $\theta^1
\theta^2 =- \theta^2 \theta^1$ and $\theta^1 \theta^1 =0$.   In
our case we shall have one $x^\mu$ (namely $t$) and one
$\theta^\a$ (which we denote by $\theta$).  In superspace one can
introduce superfields: fields which depend both on $x$ and
$\theta$.Because $\theta^2=0$, the superfields we consider can be
expanded as $\phi(t,\theta)=\varphi(t)+\theta\psi (t)$. This
concludes the three definitions.

Supersymmetric quantum field theories have remarkable properties.
Leaving aside the physical motivations for studying these
theories, they also form useful toy models.  We present here an
introduction to supersymmetric field theories with rigid and
local supersymmetry, both in $x$-space and in superspace, in the
simplest possible model%
\footnote{ Actually, an even simpler model than the one we present in these lectures
exists. It contains constant fields, so fields which do not even
depend on $t$ and $\theta$. These so-called matrix models are
important in string theory, but they do not have enough structure
for our purposes, so we do not discuss them.}. 
To avoid the complications due to ``Fierz rearrangements" 
(recoupling of four fermionic fields $A,B,C,D$ from the structure
  $(AB) (CD)$ to $(AD)(CB)$)\footnote{Here  $(AB)$ means contraction of spinor fields $A$ and $B$.} \note{*}
we consider one-component (anticommuting) spinors. Then
$(AB) (CD)$ is simply equal to $-(AD) (CB)$.  The simplest case in
which spinors have only one component is a one-dimensional
spacetime, i.e., quantum mechanics.  The corresponding superspace
has one commuting coordinate $t$ and one anticommuting coordinate
$\theta$.  Both are real.

We repeat and summarize: one can distinguish between rigidly supersymmetric field
theories,
which have a constant symmetry parameter, and
locally supersymmetric field theories whose symmetry parameter is an arbitrary
space-time dependent parameter.
For a local symmetry one needs a gauge
field.   For supersymmetry the gauge field has been called the
gravitino.  (The local symmetry on which Einstein's theory of gravitation is based is diffeomorphism invariance.  The gauge field is the metric field $g_{\mu\nu} (x)$).  Gauge theories of supersymmetry
(thus theories with a local
supersymmetry containing the gravitino) need curved spacetime.  In
other words, gravity is needed to
construct gauge theories of supersymmetry, and for that reason local
supersymmetry is usually called
supergravity.  In curved space the quanta of the metric $g_{\mu\nu}$ are massless particles 
called gravitons.  They are the bosonic partners of the gravitinos.  Neither gravitons nor gravitinos have ever been directly detected.  Classical gravitational radiation may be detected in the years ahead, but the gravitons (the quantized particles of which the gravitational field is composed) are much harder to detect individually.  The detection of a single gravitino would have far-reaching consequences.

In the last chapter we quantize the supergravity action which we
obtained in chapter~3. There are several methods of quantization,
all in principle equivalent, but we shall only discuss the BRST
method. It yields the ``quantum action", which is the action to
be used in path integrals. This method has a beautiful and
profound mathematical structure, and that is one of the reasons
we chose to include it.

The author wrote in 1976 with D.Z. Freedman and S. Ferrara the
first paper on supergravity, soon followed by a paper by S. Deser and
B. Zumino. However, we will not discuss past work
and give references; rather, the present account may serve as a
simplified introduction to that work. For readers who want to
read further, we include a few references at the end.

\mathversion{bold}
\section{Rigid $N=1$ supersymmetry in $x$-space}
\mathversion{normal}

The model we consider contains in $x$-space (or rather $t$-space)
two point particles which correspond to a real bosonic field
$\varphi (t)$ and a real fermionic field $\l(t)$.  We view them
as fields whose space-dependence (the dependence on $x$, $y$, $z$)
is suppressed. The function $\varphi (t)$ is a smooth function of
$t$, so its derivative is well-defined, but for  every value of
$t$ the expression $\l (t)$ is an independent Grassmann
number~\cite{Berezinfa}.% 
\footnote{For a more recent mathematical treatment, 
see  ``Five Lectures on Supersymmetry'' by D.~Freed (AMS)
and articles by Deligne \& Morgan and by Deligne \& Freed in 
``Quantum Fields and Strings: A course for mathematicians'' (AMS)}    \note{*}
 So $\l (t_1) \l (t_2) =- \l (t_2) \l
(t_1)$. We assume that the concept of a derivative of $\l (t)$
with respect to $t$ can be defined, and that we may partially
integrate.  As action for these ``fields" we take $S = \int L dt$
with 
\eqa L \; ({\rm rigid}) = {1 \over 2} \dot\varphi^2 + {i
\over 2} \l \dot\l \label{1one}. \eqae 
The $\dot\l = {d \over dt} \l$ 
are independent Grassmann variables, so $\dot\l (t) \l (t) =-
\l (t) \dot\l (t)$ and $\dot\l (t_1) \dot\l (t_2) =- \dot\l (t_2)
\dot\l (t_1)$.  In particular they anticommute with themselves
and with each other at equal $t$:
\eqa \{ \l (t) , \l (t) \} =0, \quad
                 \{ \l (t), \dot\l (t) \} =0, \quad
                 \{ \dot\l (t), \dot\l (t) \} =0. 
\eqae The symbol $\{ A,B \}$ is by definition $AB+BA$.  Later we
shall define Poisson brackets and Dirac brackets, which we denote
by $\{ A,B \}_P$ and $\{ A,B \}_D$ to avoid confusion.   At the
quantum level the Poisson and Dirac brackets are replaced by
commutators for commuting fields and anticommutators for
anticommuting fields, which we denote by $[A,B]$ and $\{ A,B \}$,
respectively, and which are defined by $[A,B] =AB-BA$ and $\{
A,B  \} = AB+BA$.

We introduce a concept of hermitian conjugation under which $\varphi (t)$ and $\l (t)$ are 
real: $\varphi (t)^\dagger = \varphi (t)$ and $\l (t)^\dagger = \l (t)$.  Also $\dot\l (t)$ 
is real. Furthermore $(AB)^\dagger = B^\dagger A^\dagger$ for any $A$ and $B$. We define the
action by $S=\int L(t)dt$.  
The action should be hermitian according to physical principles
(namely unitarity%
\footnote{``Unitarity'' means ``conservation of probability'': 
the total sum of the probabilities that a given system can decay into any other system should be one.}). 
 We need then a factor $i$ in the second term in (\ref{1one}) in order that $({i \over 2}
\l \dot\l )^\dagger = -{i \over 2}
\dot\l \l$ be equal to ${i \over 2} \l \dot\l$.   In the action we need $\l (t)$ at different $t$.  We repeat that for
different $t$ the $\l (t)$ are independent Grassmann variables.  Thus we need an infinite basis for all Grassmann variables.%
\footnote{ In some mathematical studies one takes a finite-dimensional basis for the Grassmann variables. 
 This is mathematically consistent, but physically unacceptable: it violates unitarity.}

Physical intermezzo which can be skipped by mathematicians:  The
term ${1 \over 2} \dot\varphi^2$ is a truncation of the
Klein-Gordon action to an $xyz$ independent field, and the term
${i \over 2} \l \dot\l$ is the truncation of the Dirac action for
a real%
\footnote{ Real spinors are called Majorana spinors,
 and complex spinors are called Dirac spinors. 
Already at this point one can anticipate that $\l$ must be real because we
took $\varphi$ to be real, and we shall soon prove that there
exists a symmetry between $\l$ and $\varphi$.} 
spinor to one of its components.  In higher dimensions the Dirac action in curved
space reads (as discussed in detail in 1929 by  H. Weyl)   \note{*}
\eqa \cl \; ({\rm Dirac}) = - ( \det e_\mu{}^m ) \bar\l
\g^m e_m{}^\mu D_\mu \l \label{1two}, \eqae where $D_\mu \l =
\del_\mu \l + {1 \over 4} \o_\mu{}^{mn} \g_{mn} \l$ with $\g_{mn}
\equiv  {1 \over 2} [ \g_m , \g_n ]$ the Lorentz generators
(constant matrices) and $\o_\mu{}^{mn}$ the spin connection (a
complicated function of the vielbein fields $e_\mu{}^m$). The
matrices $\gamma^m$ (with $m=0,1,2,\dots ,d-1$) in $d$ spacetime
dimensions satisfy  Clifford algebra relations,
$\{\gamma^m,\gamma^n\}=2\eta^{mn}$. The ``vielbein" fields
$e_\mu{}^m$ are the square root of the metric $g_{\mu\nu}$ in the
sense that $e_\mu{}^m e_\nu{}^n \eta_{mn} = g_{\mu\nu}$, where
$\eta_{mn}$ is the Lorentz metric (a diagonal matrix with
constant entries $(-1, +1, +1, \cdots , +1)$). Furthermore,
$e_m{}^\mu$ is the matrix inverse of $e_\mu{}^m$, and 
$\bar\lambda$ is defined by $\l^\dagger i \g^0$. However, in one
dimension there are no Lorentz transformations, hence in our toy
model $D_\mu \l$ is equal to $\dot\l$. Furthermore, $\det
e_\mu{}^m = e_\mu{}^m$ in one dimension, and this cancels the
factor $e_m{}^\mu$.  Thus even in curved space, the Dirac action
in our toy model reduces to ${i \over 2} \l \dot\l$
(for real $\l$; 
the factor $\frac12$ is used for real fields, just as  for
$\frac12{\dot\varphi}^2$ in~(\ref{1one})). As a consequence, the
gravitational stress tensor, which is by definition proportional
to ${\delta \over \delta e^m_\mu (x)} S$, vanishes in this model
for $\l (t)$. Also the canonical Hamiltonian %
\footnote{The momenta are defined by left-differentiation: $p=\tfrac{\partial}{\partial\dot{q}}S$. 
Hence $p(\phi)=\dot{\phi}$ and $\pi(\lambda)=-\tfrac{i}{2}\lambda$, where $\pi$ denotes the conjugate momentum of $\lambda$.}
$H= \dot{q}p-L$
vanishes for $L$ given in (2.1) and $q = \l (t)$. This will play a role in the
discussions below. The sign of the term ${1 \over 2}
\dot\varphi^2$ is positive because it represents the kinetic
energy, but the sign of the fermion term could have been chosen
to be negative instead of positive. (Requiring $\l$ to be real,
we cannot redefine $\l \rightarrow i \l$ in order to change the
sign of the second term and still keep real $\l$.)  The $+$ sign
in (2.1) will lead to the susy anticommutator $\{Q,Q\}=2H$
instead of $\{Q,Q\}=-2H$ with a hermitian $Q$.  End of physical
intermezzo.

The supersymmetry transformations should transform bosons into fermions, and vice-versa, so $\varphi$ into
$\l$, and $\l$ into $\varphi$.  Since
$\varphi$ is commuting and $\l$ anticommuting, the parameter must be
anticommuting.  We take it to be a
Grassmann number $\epsilon$, although other choices are also
possible.%
\footnote{ The author has proposed long ago with J. Schwarz to consider
$\theta$'s which satisfy a Clifford algebra, 
$\{ \theta^\a , \theta^\beta \} = \gamma^{\a\b}_\mu  x^\mu$.  
%One can then introduce not only the super-curvatures
%familiar in quantum groups but also super-torsions. 
%They satisfy a Yang-Baxter algebra (see work with Bouwknegt and McCarthy)
.}  
%\bignote{This footnote looks too special for this exposition. }
 One might then be tempted to
write down $\delta \varphi = i \epsilon \l$ and $\delta \l = \varphi
\epsilon$ (where the factor $i$ is
needed in order that $\delta \varphi$ be real, taking $\epsilon$ to
be real) but this is incorrect as
one might discover by trying to prove that the action is invariant
under these transformation rules.
There is a more fundamental reason why in particular the rule $\delta
\l = \varphi \epsilon$ is
incorrect, and that has to do with the dimensions of the fields and
$\epsilon$ as we now explain.

The dimension of an action $S \equiv \int L dt$ is zero (for
$\hbar =1$).%
\footnote{ More precisely, the dimension of $H$ and $L$ is an energy, 
and $t$ has of course the dimension of time.  In quantum mechanics the Planck constant
$\hbar \equiv h/2 \pi$ has the dimension of an energy $\times$ time
(discovered by Planck in 1900).  Since $S$ has the dimension of
an energy $\times$ time, one can define dimensionless exponents
of the action by $\exp {i \over \hbar} S$.  Such exponents appear
in path integrals.  Physicists often choose a system of units such that $\hbar =1$.}
 Hence $L = {1 \over 2} \dot\varphi^2$ should
have dimension $+1$, taking the dimension of $t$ to be $-1$ as
usual for a coordinate.  It follows that the dimension of
$\varphi$ is $-1/2$ and that of $\l$ is $0$ 
\eqa
   [\varphi ] =- 1/2 ;\quad  [\l] =0 .
\eqae
From $\delta \varphi = i \epsilon \l$ we then conclude that
$\epsilon$ has dimension $-1/2$
\eqa
\; [ \epsilon ] =- 1/2 .
\eqae
Thus $\delta \varphi = i \epsilon \l$ is dimensionally correct: $[
\delta \varphi ] =- 1/2$ and $[
\epsilon \l] =- 1/2$.  Consider now the law for $\delta \l$.  The
proposal $\delta \l = \varphi
\epsilon$ has a gap of one unit of dimension: $[\delta \l ] =0$ but
$[ \varphi \epsilon ] = - 1/2 - 1/2
=-1$.  To fill this gap we can only use a derivative (we are dealing
with massless fields so we have no
mass available).  Thus $\delta \l \sim \dot\varphi \epsilon$.  We
claim that the correct factor is minus
unity, thus
\eqa
\delta \varphi = i \epsilon \l ; \quad \delta \l = -\dot\varphi \epsilon. 
\label{1five}
\eqae
By correct we mean that (\ref{1five}) leaves the action invariant as
we now show.
It is  easy to show that $S$ (rigid) is invariant under these
transformation rules if $\epsilon$
is constant (rigid supersymmetry).  Let us for future purposes
already consider a local $\epsilon$
(meaning $\epsilon (t)$) and also keep boundary terms due to partial
integration.  One finds then if one successively varies the fields in $S$ according to (\ref{1five})
\eqa
\delta S &=& \int \left( \dot\varphi \delta \dot\varphi + {i \over 2} \delta \l  \dot\l
 +  {i \over 2} \l \delta \dot\l \right) dt \nn
&=& \int \left[ \dot\varphi {d \over dt} (i \epsilon \l) - {i
\over 2} (\dot\varphi \epsilon) \dot\l - {i \over 2} \l {d \over
dt} ( \dot\varphi \epsilon ) \right] dt \nn &=& \int \left[
\dot\varphi i \dot\epsilon \l + \dot\varphi i \epsilon \dot\l -
{i \over 2} \dot\varphi \epsilon \dot\l - {i \over 2} {d \over
dt} (\l \dot\varphi \epsilon) + {i \over 2} \dot\l \dot\varphi
\epsilon \right] dt \nn &=& \int \left[ \dot\epsilon (i
\dot\varphi \l ) - {i \over 2} {d \over dt} ( \l \dot\varphi
\epsilon ) \right] dt \label{1six} . 
\eqae 
We performed a partial
integration in the third line and used $\dot\l \epsilon =
-\epsilon \dot\l$ in the fourth line. We now assume that
``fields" (and their derivatives) tend to zero at $t= \pm
\infty$.  (If there would also be a space dimension $\s$, we
could consider a finite domain $0 \leq \s \leq \pi$, and then we
should specify boundary conditions at $\s = 0, \pi$.  This
happens in ``open string theory".)  It is clear that neglecting
boundary terms at $t= \pm \infty$, and taking $\epsilon$ constant
($\dot\epsilon =0$), the action is invariant.  (A weaker
condition which achieves the same result is to require that the
fields at $t \rightarrow + \infty$ are equal to the fields at $t
\rightarrow - \infty$).  This assumption that fields vanish at
$t= \pm \infty$ is not at all easy to justify, but we just accept
it.

The algebra of rigid supersymmetry transformations reveals that
supersymmetry is a square root of
translations, in the sense that two susy tranformations (more
precisely, a commutator) produce a
translation.  On $\varphi$ this is clear
\eqa
&& [ \delta (\epsilon_2) , \delta (\epsilon_1) ] \varphi =  
\delta (\epsilon_2) i \epsilon_1 \l - \delta (\epsilon_1) i \epsilon_2 \l 
% 1 \leftrightarrow 2 
   \nn
&& = i \epsilon_1 (-\dot\varphi \epsilon_2 )-  i \epsilon_2 (-\dot\varphi \epsilon_1 )   =
               (2 i \epsilon_2 \epsilon_1 ) \dot\varphi.
\eqae
We recall that the symbol $[A,B]$ is defined by $AB-BA$, so $[ \delta (\e_2) , \delta (\e_1) ]$ is a commutator of two supersymmetry transformations.
We used in the last step that $\epsilon_1 \epsilon_2 = - \epsilon_2 \epsilon_1$.
The result is a translation $(\dot\varphi)$ over a distance $\xi =
2i \epsilon_2 \epsilon_1$.  The same result is obtained for $\l$
\eqa
&& [ \delta (\epsilon_2 ), \delta (\epsilon_1) ] \l =
-\delta (\epsilon_2 ) \dot\varphi \epsilon_1 + \delta (\epsilon_1 ) \dot\varphi \epsilon_2
% 1 \leftrightarrow 2 =  
\nn
&& = -{d \over dt} ( i \epsilon_2 \l) \epsilon_1 + {d \over dt} ( i \epsilon_1 \l) \epsilon_2
= ( 2 i \epsilon_2 \epsilon_1 ) \dot\l.
\eqae
We used that $\epsilon_2$ is constant, so ${d \over dt} \epsilon_2 =0$, and $\l \e_1 = -  \e_1 \l$.

In higher dimensional theories this commutator on a fermion yields in addition to a translation also
a term proportional to the field equation of the fermion, and to
eliminate this extra term with the field equation, one introduces
auxiliary fields.  (Auxiliary fields are fields which appear in the action without derivatives; they are usually bosonic fields which enter as $a (t)^2$).  Here, however, the translation $\dot\l$
and the field equation of
$\l$ are both equal to $\dot\l$.  So the result could still have been a sum of the same
translation as on $\varphi$, and a field
equation, because both are proportional to $\dot\l$.  This is not the
case:  the coefficient of $\dot\l$ is the same as the coefficient
of $\dot\varphi$.   There is a simple counting argument that explains
this and that shows that no auxiliary
fields are needed in this model. Off-shell (by which physicists mean: when the field equations are not satisfied) the translation operator
is invertible (the kernel of ${\del \over \del t}$ with the boundary conditions mentioned before is empty), hence the commutator
$[\delta (\epsilon_2), \delta (\epsilon_1)]$ cannot vanish on field
components.  It follows that under rigid
supersymmetry if  ``the algebra closes" (meaning if $[ \delta (\epsilon_2 ),
\delta (\epsilon_1 )]$ is uniformly equal to only a translation but
no further field equations), each bosonic field
component must be mapped into a fermionic
one, and vice-versa.  Then the number of bosonic field components
must be equal to the number of
fermionic field equations.  In our toy model there is one bosonic
field component $(\varphi)$ and one
fermionic field component $(\l)$.  Thus there are no auxiliary fields
needed in this model.%
 \footnote{ In the 4-dimensional Wess-Zumino model there
are 2 propagating real scalars (A and B) and
a real 4-component spinor. Hence there one needs two real bosonic
auxiliary fields (F and G). In the 2-dimensional heterotic string the
right-handed spinors $\l_R$ do not transform under rigid supersymmetry, $Q \l_R =0$. 
 It follows that on the right-hand side of the susy commutator evaluated on
$\l_R$ the field equation $(\dot\l)$ exactly cancels the translation $P \l = \dot\l$.}

We can construct charges $Q$ and $H$ which produce susy and
time-translation transformations.  This requires
equal-time Poisson brackets for $\varphi$, and Dirac brackets for
$\l$, which become at the quantum level
commutators and anticommutators. For $\varphi$ these results are
standard: the conjugate momentum $p$ of $\varphi$ is defined by $p = {\partial \over \partial \dot\varphi} S$ and this yields $p = \dot\varphi$.  The quantum commutator is given by
\eqa
p = \dot\varphi;\quad [p (t) , \varphi (t) ] = {\hbar \over i}.
\eqae
For $\l$ the conjugate momentum is (we use left-derivatives) $\pi =
{\del \over \del \dot\l} S = - {i
\over 2} \l$.  The relation $\pi =- {i \over 2} \l$ is a constraint between the coordinates and the conjugate momenta, called by Dirac a primary constraint
\eqa
\Phi = \pi + {i \over 2} \l =0.
\eqae
The naive Hamiltonian is $H_L = \dot{Q} P -L = \dot{q} p- {1 \over 2}
\dot{q}^2 + \dot\l \pi - {i \over 2}
\l \dot\l = {1 \over 2} p^2$.  Here $Q$ and $P$ denote the total set of fields and their canonically conjugates.  (We must put $\dot \l$ in front of
 $\pi$, if we define $\pi$ by left-differentiation,
 $\pi = {\del \over \del \dot\l} L$, because only then
$H_L$ is independent of $\dot{q}$ and $\dot \l$. Namely $\delta
H_L$ contains no terms with $\delta \dot{Q}$ but only with
$\delta Q$ and $\delta P$.) According to Dirac, one must then
consider the naive Hamiltonian plus all possible primary
constraints \eqa H = {1 \over 2} p^2 + \a \left( \pi + {i \over
2} \l \right), \eqae where $\a (t)$ is an arbitrary anticommuting
parameter.  Requiring that the constraint $\pi + {i \over 2} \l =
0$ be maintained in time requires $[H, \pi + {i \over 2} \l ] =0$
modulo the constraints, which is indicated by the symbol~$\approx$
\eqa \left[ H, \pi + {i \over 2} \l \right]_P \approx 0. \eqae The
subscript $P$ indicates that we use here Poisson brackets. We
define the Poisson bracket by \eqa \{ f (p, q) , g (p , q) \}_P =
- \partial f / \partial p {\partial \over \partial q} g + (-)^\s
\partial g / \partial p {\partial \over \partial q} f, \eqae where
$\s = + 1$ except when both $f$ and $g$ are anticommuting, in
which case $\s =- 1$.  The basic  relations are $\{ p , q \}_P
=-1$ and $\{\pi,\l\}_P =-1$. Of course ${1 \over 2} p^2$ commutes
with $\pi+ {i \over 2} \l$, but \eqa \left\{ \Phi, \Phi
\right\}_P = \left\{ \pi + {i \over 2} \l, \pi + {i \over 2} \l
\right\}_P = - {i \over 2} - {i \over 2} =- i. \eqae Note that the
Poisson bracket $\left\{ p, q \right\}_P$ is $-1$ for bosons and
fermions alike.%
\footnote{ In quantum mechanics the sign of the quantum commutator $[p,q] = - i \hbar$ or the
quantum anticommutator $\{ \pi , \l \} = - i \hbar$ is not a
matter of convention but follows from the compatibility of the
field equations with the Heisenberg equations.  For example, for a
Dirac spinor $\psi$ with mass $m$ one has $L=i\psi^\dagger
\dot\psi+ m\psi^\dagger \psi$ and $\pi=-i\psi^\dagger$. For
$\psi$ the field equation is $i\dot\psi+m\psi=0$, and the
Heisenberg equation is $\dot\psi={i \over \hbar} [H,\psi]$ with
$H=-m\psi^\dagger \psi$. Compatibility of the field equation with
the Heisenberg equation requires
$\{\psi,\pi\}=\{\psi,-i\psi^\dagger\}=-i\hbar$ which agrees with
$\{\pi,\psi\}={\hbar \over i}$ in quantum brackets.}  It follows
that $[H, \pi + {i \over 2} \l ] = - i \a$, and hence $\a =0$.
Thus, with $\a =0$, there are no further (secondary) constraints,
and we have \eqa H= {1 \over 2} p^2. \eqae

Whenever a set of constraints $\phi^\a$ satisfies $\{ \phi^\a ,
\phi^\beta \} = M^{\a\b}$ with ${\rm sdet} M^{\a\b} \not= 0$, we
call these constraints second class
constraints.%
\footnote{  The expression ${\rm sdet} M$ denotes the superdeterminant of a supermatrix 
$M = \left( \begin{array}{cc} A & B \\ C & D \end{array} \right)$,
where $A$ and $D$ contain commuting entries and $B$ and $C$
anticommuting entries.  Any matrix can be written as the product
of diagonal matrices 
$\left( \begin{array}{cc} A & 0 \\ 0 & B
\end{array} \right)$ and triangular matrices $\left(
\begin{array}{cc} I & C \\ 0 & I  \end{array} \right)$ and $\left(
\begin{array}{cc} I & 0 \\ D & I  \end{array} \right)$. Namely, $\left(
\begin{array}{cc} A & B \\ C & D  \end{array} \right)=\left(
\begin{array}{cc} I & BD^{-1} \\ 0 & I  \end{array} \right) \left(
\begin{array}{cc} A-BD^{-1}C & 0 \\ 0 & D  \end{array} \right)\left(
\begin{array}{cc} I & 0 \\ D^{-1}C & I  \end{array} \right)$.  The
superdeterminant of the product of supermatrices is the product
of the superdeterminants of these supermatrices, and ${\rm sdet} \left(
\begin{array}{cc} A & 0 \\ 0 & D \end{array} \right) = \det A/
\det D$ while ${\rm sdet} \left( \begin{array}{cc} I & B \\ 0 & I
\end{array} \right) =1$. Hence ${\rm sdet} M =\det (A-BD^{-1}C)/\det D $. }
It follows that $\phi = \pi + {i \over
2} \l$ is a second class constraint. The Dirac bracket is
defined by \eqa \left\{ A,B \right\}_D = \left\{ A,B \right\}_P -
\left\{ A, \Phi \right\}_P \left\{ \Phi, \Phi \right\}^{-1}_P
\left\{ \Phi, B \right\}_P \label{1bracket}, \eqae where $\{ A,B
\}_P$ denotes the Poisson bracket. Its definition is chosen such
that $\{A,\Phi\}_D=0$ for any $A$ and any second-class constraint
$\Phi$.  Since in our toy model $\{ \Phi, \Phi \} =-i$, we find
\eqa \{ A,B \}_D = \{ A,B \}_P - i \{ A, \pi+ {i \over 2} \l \}_P
\{ \pi + {i \over 2} \l, B \}_P \label{1sixteen}. \eqae

We can now compute the basic equal-time Dirac brackets
\eqa
&& \{ \l (t), \l (t) \}_D =0 - i (-1) (-1) =-i, \nn
&& \{ \pi (t) , \l (t) \}_D = -1 - i \left( {-i \over 2} \right) (-1)
=- {1 \over 2}, \nn
&& \{  \pi (t), \pi (t) \}_D =0 - i \left( {-i \over 2} \right) \left(
{-i \over 2} \right)  = {i \over 4}.
\eqae
Recalling that $\pi =- {i \over 2} \l$, we see that these relations
are consistent: we may replace $\pi$ by $-{i \over 2} \l$ on the
left-hand side.  At the quantum level, as first proposed by Dirac, we add a factor $i \hbar$ to
the Poisson brackets to obtain the quantum (anti) commutators.  Hence
\begin{align} \label{1varphi} \{ \l (t), \l (t) \} & = \hbar,\nonumber \\
[ p (t) , \varphi (t) ] & = {\hbar \over i}.
\end{align}

We now construct the susy charge $Q$ as a Noether charge.  A Noether charge is the space integral of the time component of the Noether current, but since there is no space in our toy model, the Noether current is equal to the Noether charge.  We want to obtain an expression for the Noether charge in terms of $p$'s and $q$'s, and therefore we rewrite (\ref{1one}) in Hamiltonian form, namely as $L = \dot{q} p-H$ where $H$ depends only on $p, \pi, \varphi, \l$ but not on their time derivatives.  Since, as we shall discuss, the terms proportional to a derivative of the symmetry parameter yield the Noether current, the latter will only be a function of $p$'s and $q$'s but not of derivatives of $p$'s and $q$'s.

The action in Hamiltonian form reads \eqa\label{achamil} L=
\dot\varphi p+ \dot\l \pi - {1 \over 2} p^2. \eqae where we took
the Dirac Hamiltonian $H = {1 \over 2} p^2$ as discussed above.
This action is invariant under 
\begin{gather} \label{1twenty} \delta \varphi = {i \over 2}
\epsilon \l - \epsilon \pi, \quad \delta \l = -p \epsilon, \nonumber \\
\delta p =0, \quad \delta \pi = {i \over 2} p \epsilon.
\end{gather}
These rules follow by requiring
invariance of the action, but one can also derive them by adding
equation of motion symmetries to the original rules. For example,
$p=\dot \varphi$ leads to $\delta p={i \over 2}\epsilon
\dot\l-\epsilon\dot\pi$, and to remove $\dot\l$ one may add
$\delta({\mathrm extra}) p={i \over 2}\epsilon{\del \over \del \pi}S$
and $\delta({\mathrm extra}) \pi=-{i \over 2}{\delta S \over \delta
p} \epsilon$.  These extra transformation rules form a separate
symmetry of any action, so we may add them to the original
rules.  Then $\delta p=-\epsilon \dot\pi$, and also $\delta
\pi=\delta (-{i \over 2}\l)={i \over 2}\dot\varphi \epsilon$ is
modified into $\delta \pi={i \over 2}\dot\varphi \epsilon-{i
\over 2}(\dot\varphi-p)\epsilon={i \over 2}p\epsilon$. To also
 remove the term
$-\epsilon\dot\pi$ in $\delta p$, one adds another equation of motion symmetry:
$\delta(\mathrm{extra})p = {\delta \over \delta \l} S \e$ and
$\delta(\mathrm{extra})\l=\tfrac{\partial}{\partial p}S$. The final result
is~(2.22).

The standard way of obtaining the Noether charge follows from
letting the rigid parameter become local and collecting terms
proportional to $\dot\e$.   The terms with $\dot\epsilon$ in
$\delta L$ for local $\epsilon (t)$ are contained in \eqa \delta
L &=& p {d \over dt} \left(  {i \epsilon \over 2} \l - \epsilon
\pi \right) - \pi {d \over dt} (-p \epsilon ) + \dot\l \left( { i
\over 2} p \epsilon \right) \nn &=& {i \over 2} p \dot\epsilon \l
- p \dot\epsilon \pi + {d \over dt} ( \pi p \epsilon).
\label{1twentyone} \eqae So, defining $Q$ as the coefficient of
$i\dot \epsilon$, we find \eqa Q = p \left( {1 \over 2} \l +i \pi
\right). \label{1twentytwo} \eqae This reproduces (2.22);
for example, $\tfrac{1}{\hbar}[\varphi, \epsilon Q]=\delta \varphi$ follows
from~(\ref{1varphi}). Using the constraint $\pi =-{i \over 2} \l$
we see that $Q$ becomes equal to $Q = p \l$, and (2.22)
reduces to (\ref{1five}). However, if one uses (anti) commutators
one needs a Hamiltonian treatment, and then one needs
(\ref{1twentytwo}).

The charge $Q$ which appears in brackets such as $[ \varphi, \e Q
]$ is clearly an operator, so $Q$ in (\ref{1twentytwo})  is an
operator expressed in terms of Heisenberg fields.  The latter
satisfy their own equations of motion.  On the other hand, in the
action the fields are off-shell.  So, in principle one might need
extra terms proportional to the equations of motion to obtain the
correct off-shell transformations.  In this case we do not need
such terms.  In section 6 we shall discuss the Hamiltonian
approach with off-shell fields.  This is a very general approach
which yields the action in Hamiltonian form and the quantum BRST
charge, starting only from the set of first class constraints.

The other way of obtaining $Q$ (more precisely, $i\epsilon Q$) is to write
it as a sum of terms of
the form $\delta \varphi p$ for all fields, plus $-K$ where
$\delta L = {d \over dt} K$.  From ({\ref{1twentyone}) we read off
that $K =  \pi p \epsilon$.  Hence
\begin{align}
i \epsilon Q  & = \delta \varphi p + \delta \l \pi - K = \left( {i
\over 2} \epsilon \l - \epsilon \pi
\right) p - p \epsilon \pi - \pi p \epsilon \nonumber \\
& = i \epsilon \left( {1 \over 2} \l + i \pi \right) p,
\end{align}
which is indeed the same result as in (\ref{1twentytwo}).

The supersymmetry algebra (rather a superalgebra with commutators
and anticommutators) is now easy to evaluate.  Using the quantum
brackets of (\ref{1varphi}) finds \eqa && \{  Q,  Q \} = \left\{
\left( { 1\over 2} \l + i\pi \right) p , \left( {1\over 2} \l +
i\pi \right) p \right\} = \{  \l p,  \l p \} =\hbar p^2 =2\hbar H,
\nn && [H, Q] =0 \quad \text{(via Jacobi, or directly)}.
\label{Jacobi} \eqae Thus the generators $Q$ and $H$ form a
closed superalgebra, and supersymmetry is the square root of (the
generator of time-) translations.

Dirac was the first to take the square root of the Laplace
operator $\Box$, and this led to the famous Dirac equation of
1927.  This equation led to the prediction that for  every
fermionic particle there is a fermionic antiparticle. These
antiparticles have been found in the laboratories. Likewise, the
square root $Q\sim \sqrt{H}$ predicts that for every particle there
should be a superpartner.  Not a single superpartner has been
found so far, but that may change.

\mathversion{bold}
\section{$N=1$ supergravity in $x$-space}
\mathversion{normal}

Having discussed the rigid supersymmetry (= susy) of the action
$S  \; {\rm (rigid)} \;  = \int L dt$ with $L=  {1 \over 2}
\dot\varphi^2 + {i \over 2} \l \dot\l$, we turn to local susy. We
let $\epsilon$ become time-dependent and find then (see
(\ref{1six})). \eqa \delta S \; {(\rm rigid)} \; =
\int\limits^\infty_{-\infty} \dot\epsilon (i \dot\varphi \l) dt.
\label{1seee} \eqae The boundary terms at $t = \pm \infty$ vanish
if we require that $\e (t)$ vanishes at $t = \pm \infty$. To
cancel this variation we introduce the gauge field for local
susy, the gravitino $\psi (t)$.  The transformation rule of a
gauge field begins always with a derivative of the local
parameter.  We then couple $\psi$ to the Noether current of rigid
supersymmetry, using $\delta \psi = \dot\epsilon + \cdots$ to fix
the overall constant of this new term \eqa S \; {\rm (Noether)}
\; = \int\limits^\infty_{-\infty}  (- i \psi \dot\varphi \l) dt ;
\delta \psi = \dot\epsilon + \cdots \label{1already} \eqae If we
vary $\psi$ in (\ref{1already}), then the variation $\delta S$
(Noether) cancels $\delta S$ (rigid), but the fields
$\dot\varphi$ and $\l$ in $S$ (Noether) must also be varied. This
yields two further variations \eqa \delta S \; {\rm (Noether)} \;
= \int\limits^\infty_{-\infty} \left[ - i \psi \left\{ {d \over
dt} (i \epsilon \l ) \right\} \l + i \psi \dot\varphi \epsilon
\dot\varphi \right] dt. \eqae
%Since $\l \l =0$, the derivative ${d \over dt}$ must cut on $\epsilon$.  Then
In the first variation the ${d \over dt}$ must hit the field $\l$
because otherwise one would be left with $\l\l$ which vanishes.
Hence the remaining variations to be canceled are \eqa \delta S
\; {\rm (Noether)} \; = \int\limits^\infty_{-\infty}  i \psi
(\dot\varphi \dot\varphi + i \l \dot\l ) \epsilon dt.
\label{1twentyseven} \eqae The last term can be canceled by
adding a new term in $\delta \l$ (because this variation is
porportional to the field equation of $\l$). However, the first
term can only be canceled by introducing a new field $h$ (the
graviton) and coupling it to $\dot\varphi \dot\varphi$.  Thus the
coupling of rigidly supersymmetric matter to the supergravity
gauge fields requires for consistency (invariance of the whole
action under local susy) also the coupling to gravity.  \emph{
Local susy is a theory of gravity, and this explains the name
supergravity.}

There appears, however, an ambiguity at this point:  we can also
couple this new field $h$ to $-i \l \dot\l$, and the most general
case is a linear combinations of both possibilities.  Hence we add
\begin{align} \label{1however}  S\,{\rm (stress)} & = - \int\limits^\infty_{-\infty} h
[ \dot\varphi \dot\varphi - i \l \dot\l x ] dt, \nonumber \\ \delta h & =
-i \epsilon \psi, \quad \delta \l =- \dot\varphi \epsilon +i (1 +x)
\psi \l \epsilon, \end{align} where $x$ is a free
constant real parameter. We must now evaluate the old variations
in the new action,  the new variations in the old action, and the
new variations in the new action. The aim is to use these
variations to cancel~(\ref{1twentyseven}).

The new variation $\delta h =- i \epsilon \psi$ in the new action $- \int h \dot\varphi
\dot\varphi dt$ cancels the first term in
(\ref{1twentyseven}).  The new variation $\delta \l =  i (1+x) \psi \l
\epsilon$ in the old action $ {i \over 2} \l \dot\l$ in (\ref{1one}) and the
new variation $\delta h =-i \epsilon \psi$ in the new action $S$ (stress) cancel the
variation $- \psi \l \dot\l \epsilon$ of $\delta S$ (Noether)
\begin{gather} i \l {d \over dt} [ i (1+x) \psi \l \epsilon ] - i \epsilon \psi
(i \l \dot\l x) - \psi \l \dot\l \epsilon   \nonumber \\
 = (1 +x) \l \dot\l \psi \epsilon - \psi \epsilon \l \dot\l x- \psi
\epsilon \l \dot\l = 0.
\end{gather}
(We partially integrated the first term, and the last term comes from (\ref{1twentyseven})).
We find thus at this moment a free parameter $x$ in the action and
transformation rules; this frequently
happens in the construction of supergravity models, and usually these
parameters get fixed at a later stage, or they can be removed by field redefinitions.  We demonstrate this later explicitly in our toy model.

We are again in the same situation as before: we canceled $\delta
S$ (Noether) by introducing a new term in the action, namely $S$
(stress).   We already took into account the variation of the
gauge field $h$ in this new term, but we must still vary the
matter fields in $S$ (stress).  We first set $x=0$ and later
consider the case $x \not= 0$.  If $x=0$, we need only vary the
$\dot\varphi$ in $S$ (stress) and this yields 
\begin{align}  \label{1streee} \delta L\,
{\rm (stress)} = - 2h \dot\varphi {d \over dt} (i \epsilon \l) = -2 h \dot\varphi i \dot\epsilon \l -2 h \dot\varphi i
\epsilon \dot\l.
 \end{align} The first term is
proportional to the Noether current $\dot\varphi \l$ in
(3.2) and can thus be canceled by a new term in the
gravitino law \eqa \delta \; {\rm (new)} \; \psi =- 2 h
\dot\epsilon. \eqae Substituting this vatiation into
(\ref{1already}), the first term in (\ref{1streee}) is
cancelled.   The second term in (\ref{1streee}) is proportional
to the free field equation of $\l$ and can be canceled by adding
a new term to the transformation law of $\l , \delta \l = 2 h
\dot\varphi \epsilon$, because then $\delta ({i \over 2}\l
\dot\l)=-i\dot\l (2h\dot\varphi \epsilon)$ cancels the second
term in (\ref{1streee}).

The new transformation law $\delta \l = 2 h \dot\varphi \epsilon$
produces a new variation in the Noether action (\ref{1already})
\eqa
\delta L\,(\text{Noether due to} \; \delta \l = 2 h
\dot\varphi \epsilon) =- i \psi \dot\varphi (2 h \dot\varphi
\epsilon).
\label{1Noether}
\eqae
Since this term is proportional to $\dot\varphi \dot\varphi$ it can
be canceled by a final extra term in
$\delta h$, namely $\delta h = 2i h \epsilon \psi$. Then the new variation of
$h$ used in (\ref{1however}) cancels (\ref{1Noether}).

Because each time when we replace an $\dot\epsilon$ in a variation by $\psi$ we loose a time derivative,
 this process of adding further terms to the action and transformation laws is guaranteed to stop. 
 Of course it is not guaranteed that an invariant action exists. 
 Examples are known where this process does not yield an invariant action, for example adding a cosmological constant to supergravity in $10 + 1$ dimensions.

We have now canceled all variations for the case $x=0$, hence we have
constructed a locally susy action.
The final results read
\begin{align} \label{1thirtythree}
L & = {1 \over 2} \dot\varphi^2 + {i \over 2} \l \dot\l - i \psi
\dot\varphi \l - h
\dot\varphi^2, \nonumber \\
\delta \varphi & = i \epsilon \l, \quad \delta \l = -\dot\varphi \epsilon
+ i \psi \l \epsilon +2 h
\dot\varphi \epsilon, \nonumber\\
\delta \psi & = \dot\epsilon - 2 h \dot\epsilon,\quad \delta h = -i
\epsilon \psi +2 i h \epsilon \psi.
\end{align}

Before going on, we make three comments.

\medskip
\noindent 1) There is no gauge action for gravity or local
supersymmetry in one dimension as one might expect, since the scalar
curvature $R$ and its linearization, the Fierz-Pauli action,%
\footnote{The linearized form of the Einstein-Hilbert action $e R$ is called the
Fierz-Pauli action and is given in n-dimensions by
\[
L = - {1 \over 2} \varphi^2_{\mu\nu , \l} + \varphi_\mu^2 -
\varphi_\mu \varphi_{, \mu} + {1 \over 2} \varphi_{, \mu}^2,
\]
where $\varphi_\mu = \del^\nu \varphi_{\mu\nu}$, $\varphi=\eta^{\mu\nu}
\varphi_{\mu\nu}$ and the metric $g_{\mu\nu} \equiv \eta_{\mu\nu} +
\kappa h_{\mu\nu}$ is related to $\varphi_{\mu\nu}$ by
$(\sqrt{-g} g^{\mu\nu})_{\rm lin} - \eta^{\mu\nu} =- \kappa
(h^{\mu\nu}- {1 \over 2} \eta^{\mu\nu} h)=\varphi^{\mu\nu}$.  
In one dimension the first term in $L$ cancels the last term, and the second term cancels the third term.} 
vanishes in one dimension.  (Also the gravitino gauge action vanishes in one and $1+1$ dimensions. 
 The gravitational action is a total derivative in $1+1$ dimensions, where it yields the Euler invariant).  A gauge action for supergravity in one dimension
would have to start with $L = {1 \over 2} \dot{h} \dot{h} + {i \over 2}
\psi \dot\psi$ and it is indeed
invariant under the rigid symmetries $\delta h = i
\epsilon \psi, \delta \psi = -\dot{h} \epsilon$, see (\ref{1one}) and
(\ref{1five}).  However, for local $\epsilon (t)$  the
rules were already fixed by the matter coupling, see
(\ref{1thirtythree}), and these rules do not leave this action
invariant.

\medskip
\noindent 2) The term $\int g \varphi^n dt$ with $g$ a coupling
constant cannot be made supersymmetric.  Yukawa couplings
do not exist in this model because $\l \varphi \l$ vanishes.
However, one can make $\l$ a complex Dirac spinor and then
supersymmetric interactions exist. One can also supersymmetrize a term $f
( \varphi ) \dot\varphi
\dot\varphi$; the action becomes $f (\varphi ) ( \dot\varphi
\dot\varphi + i \l \dot\l)$ and is called a
susy nonlinear $\s$ model because $f(\varphi)$ can be nonlinear, for
example $\exp \varphi$.

\medskip
\noindent 3) One can also couple the first-order action
in~(\ref{achamil}) to supergravity. Denoting the graviton and
gravitino fields by $H$ and $\Psi$, the result is \eqa &&
L=\dot\varphi p + \dot\l \pi  - {1 \over 2}p^2 - i \Psi \left( {1
\over 2}p\l + i p \pi \right) - Hp^2,  \nn && \delta p=0,\quad \delta
\pi={i \over 2} p \epsilon,\quad \delta \l=-p \epsilon, \nn && \delta
\varphi={i \over 2}\epsilon \l - \epsilon \pi,\quad \delta
\Psi=\dot\epsilon,\quad \delta H=-i\epsilon \Psi. \label{1Gaussian}
\eqae Note that $\delta (\pi+{i \over 2}\l)=0$ agrees with the
constraint $\pi+{i \over 2}\l=0$. So we may replace $\pi$ by $-{i
\over 2} \l$ in the action and transformation rules. Furthermore
we can eliminate $p$ by integrating in the path integral over a
Gaussian with $p^2$ \cite{rfeynman}. The result of these
manipulations is the following action \eqa L = {1 \over 2} {1
\over 1+2H} \dot\varphi^2 - {i \over 1+2H} \dot\varphi \Psi \l +
{i \over 2} \l \dot\l. \eqae Comparison with the second-order
action in (\ref{1thirtythree}) (the action without conjugate
momenta) we can read off how $H$ is related to $h$, and $\Psi$ to
$\psi$. \eqa\label{dvatr} H= {h \over 1-2h} ,  \;\; \Psi \; = {1
\over 1-2h} \psi  , \;\; {\rm or} \;\; h = {H \over 1+2H}  , \;\;  \psi
\; = {1 \over 1+2H} \Psi. \eqae The Jacobian for the change of
variables from $(H,\Psi)$ to $(h,\psi)$ is ${1 \over
1-2h}=1+2H$.  (One needs a super Jacobian, in particular $\del
\delta \psi / \del \psi$ is equal to $1-2 h$, and not simply
equal to $(1-2 h)^{-1}$).  The transformation rules in
(\ref{1Gaussian}) go over into (\ref{1thirtythree}) if one uses
these redefinitions.

For physicists:  The Jacobian $J = 1/1-2h$ can be exponentiated using a new kind of ghosts, introduced by Bastianelli and the author and playing the same role as the Faddeev-Popov ghosts.  In order that the theories with $h$ and $\psi$ and $H$ and $\Psi$ are equivalent at the quantum level (by which we mean that they should give the same Feynman graphs) one needs those new ghosts.  The propagators of $h$ and $\psi$ come from the gauge fixing terms.

We now return to the model in (\ref{1thirtythree}) and evaluate the local susy algebra.
On $\varphi$ one finds
\eqa
[ \delta (\epsilon_2), \delta (\epsilon_1) ] \varphi &=& 
  i \epsilon_1 (-\dot\varphi \epsilon_2 + i \psi \l \epsilon_2 + 2 h \dot\varphi \epsilon_2) - 
  i \epsilon_2 (-\dot\varphi \epsilon_1 + i \psi \l \epsilon_1 + 2 h \dot\varphi \epsilon_1)
% 1 \leftrightarrow 2 
       \nn
&=& [ 2 i (1-2 h) \epsilon_2 \epsilon_1 ] \dot\varphi + i [-2 i
\epsilon_2 \epsilon_1 \psi ] \l.
\eqae
The right-hand side contains a general coordinate transformation
$\delta \varphi = \hat\xi \dot\varphi$ with
$\hat\xi = 2i (1-2h) \epsilon_2 \epsilon_1$; this is clearly the
gravitational extension of the
nongravitational rigid translation with parameter $\xi = 2i \epsilon_2 \epsilon_1$ which we
found in the rigid susy commutator.
The second term is a local susy transformation $i \hat{\epsilon} \l$ of $\varphi$ with
parameter $\hat\epsilon = - 2 i \epsilon_2 \epsilon_1 \psi$.  Note that the composite parameters $\hat\xi$ and $\hat\epsilon$ are field- dependent.  The structure constants are no longer constants!   This has led to a new development in group theory.

On $\l$ one finds after somewhat lengthy algebra
\eqa
[ \delta ( \epsilon_2 ), \delta (\epsilon_1) ] \l = \hat\xi \dot\l -
\dot\varphi \hat\epsilon +2 h
\dot\varphi \hat\epsilon.
\eqae
and the terms with $\hat\epsilon$ agree with (\ref{1thirtythree}) (because $i \psi \l \hat\epsilon =0$ due to $\psi \psi =0$).
The term with $\hat\xi$ constitutes a general coordinate transformation on $\l$.
Clearly, the same algebra is found on $\l$ as on $\psi$!

On $\psi$ one finds
\eqa
[ \delta (\epsilon_2), \delta (\epsilon_1) ] \psi &=& 
- 2 (-i \epsilon_2 \psi + 2i h \epsilon_2 \psi ) \dot\epsilon_1 + 2 (-i \epsilon_1 \psi + 2i h \epsilon_1 \psi )\dot\epsilon_2
%- 1 \leftrightarrow 2 
\nn
&=&  -2i \left[ {d \over dt} (\epsilon_2 \epsilon_1 ) \right] \psi +
4 i h \left[ {d \over dt}
(\epsilon_2 \epsilon_1 ) \right] \psi \nn
&=& {d \over dt}  \hat\epsilon + 2i  \epsilon_2 \epsilon_1 \dot\psi
-2h  {d \over dt}
  \hat\epsilon - 4 ih \epsilon_2 \epsilon_1 \dot\psi \nn
&=& {d \over dt} \hat\epsilon - 2h {d \over dt} \hat\epsilon +
\hat\xi \dot\psi.
\eqae
So also on $\psi$ the same local algebra is realized.

Finally also on $h$ the same algebra is realized%
\footnote{Here and below ``$1 \leftrightarrow 2$'' means ``switch indices 1 and 2 in the preceding expression.''}
\eqa
&& [ \delta ( \epsilon_2), \delta (\epsilon_1)] h = 
- i \epsilon_1 ( \dot\epsilon_2 - 2h \dot\epsilon_2 ) - 2i [ \delta (\epsilon_2) h \psi ] \epsilon_1
% + (i \epsilon_2 ( \dot\epsilon_1 - 2h \dot\epsilon_1 ) - 2i [ \delta (\epsilon_1) h \psi ] \epsilon_2)
   - (1 \leftrightarrow 2)  \nn
&& =  - i \epsilon_1 (\dot\epsilon_2 - 2h \dot\epsilon_2) - 2i (-i \epsilon_2 \psi +2 i h \epsilon_2 \psi) \psi \epsilon_1 - 2i h
       (\dot\epsilon_2 - 2h \dot\epsilon_2) \epsilon_1 
% && +   ( i \epsilon_2 (\dot\epsilon_1 - 2h \dot\epsilon_1) - 2i (i \epsilon_1 \psi +2 i h \epsilon_1 \psi) \psi \epsilon_2 - 2i h
% (\dot\epsilon_1 - 2h \dot\epsilon_1) \epsilon_2 )
   - ( 1 \leftrightarrow 2 ) 
\nn
&& = - i \hat\epsilon \psi +2i h \hat\epsilon \psi + \hat\xi \dot{h} -
\dot{\hat\xi} h + {1 \over 2} \dot{\hat\xi}.
\label{1thirtyseven}
\eqae
The terms with $\hat\epsilon$ clearly agree with (\ref{1thirtythree}).
The terms with $\hat\xi$ in (\ref{1thirtyseven}) yield a general
coordinate transformation of $h$ as we shall
discuss below.  Hence, the local susy algebra closes on all fields uniformly.
We can write this as
\begin{equation}\label{moya}
[\delta_s(\epsilon_2),\delta_s(\epsilon_1)]=\delta_{s}(-2i\epsilon_2
\epsilon_1\psi)+\delta_g((1-2h)2i\epsilon_2\epsilon_1).
\end{equation}
The commutator of $\delta_s$ with $\delta_g$, and with itself,
close (they are proportional to $\delta_s$ and $\delta_g$).

We can understand the closure of the local supersymmetry algebra by using
the same argument as used for the rigid supersymmetry algebra. There is
one bosonic gauge field component ($h$) and one fermionic gauge
field component ($\psi$), hence no auxiliary fields in the gauge sector are needed.

We now consider the case $x \not= 0$.  Here a simple argument
suffices.  Rescaling \eqa \l = ( 1 +2 h x)^{1/2} \tilde\l,
\label{1thirtyeight} \eqae we obtain as action from (3.10) \eqa \cl = {1
\over 2} \dot\varphi^2 + {i \over 2} (1 + 2 hx) \tilde\l {d \over
dt} \tilde\l -i \tilde\psi \dot\varphi \tilde\l - h \dot\varphi^2,
\label{action} \eqae where we also rescaled $\psi$ according to
\eqa \psi (1+2 hx)^{1/2} = \tilde\psi. \label{1forty} \eqae We
have produced the action with $x \not= 0$ from the action with
$x=0$ by a simple rescaling of the fields $\l$ and $\psi$.  It
follows that this action is also locally susy. The precise susy
transformation rules follow from this rescaling \eqa && \delta \l
= (1+2 hx)^{1/2} \delta \tilde\l + (1+2 hx)^{-1/2} x \delta h
\tilde\l \nn &&\,\,\,\,\,\,\,\, = - \dot\varphi \epsilon + i
\tilde\psi \tilde\l \epsilon +2 h \dot\varphi \epsilon.
\label{1fortyone} \eqae By dividing (\ref{1fortyone}) by $(1+2
hx)^{1/2}$ and using $\delta h = -i \epsilon \psi +2i h \epsilon
\psi$ we find $\delta \tilde\l$. It reads \eqa && \delta \tilde\l
=- \dot\varphi \tilde\epsilon + i \tilde\psi \tilde\l
\tilde\epsilon +2 h \dot\varphi \tilde\epsilon  - x (1 + 2
hx)^{-1} (-i \tilde\epsilon \tilde\psi +2i h \tilde\epsilon
\tilde\psi ) \tilde\l, \eqae where we also rescaled $\epsilon$
according to $(1+2 hx)^{-1/2} \epsilon = \tilde\epsilon$.  (We
used $\epsilon \psi = \tilde\epsilon \tilde\psi$ in $\delta h$).
By rearranging these terms as \eqa \delta \tilde\l &=& -(1-2h)
\dot\varphi \tilde\epsilon +i \left( 1+ {x-2hx \over 1+2 hx}
\right) \tilde\psi \tilde\l \tilde\epsilon \nn &= & -(1-2h)
\dot\varphi \tilde\epsilon +i \left( {1 +x \over 1+2 hx} \right)
\tilde\psi \tilde\l \tilde\epsilon, \eqae we find a polynomial
result for $x=-1$ \eqa \delta \tilde\l = -(1-2h) \dot\varphi
\tilde\epsilon,\quad \delta \varphi =i \tilde\epsilon (1-2h) \tilde\l
\label{delta}. \eqae For $\delta h$ and $\delta \tilde\psi$ we
find in a similar manner \eqa \delta h &=& -i \tilde\epsilon
\tilde\psi +2i h \tilde\epsilon \tilde\psi, \delta \tilde\psi =
\sqrt{(1-2h)} \delta \psi = \sqrt{(1-2 h)} (1-2h) \dot\epsilon \nn
&=& (1-2 h) [(1-2 h) \dot{\tilde\epsilon} - \dot{h}
\tilde\epsilon ]. \label{polynomial} \eqae We used that $\delta h
\psi$ vanishes.  Thus for $x=-1$ all laws become again polynomial.

We can, in fact, apply the Noether method also directly to gravity.
This will explain the $\hat\xi$
terms in (\ref{1thirtyseven}).  Starting with
\eqa
L \; {\rm (rigid)} \; = {1 \over 2} \dot\varphi^2 + {i \over 2} \l \dot\l,
\eqae
we find from the translation symmetry rules $\delta \varphi = \xi
\dot\varphi$ and $\delta \l = \xi
\dot\l$ for local $\xi$
\eqa
\delta S \; {\rm (rigid)} \; = \int \left[ \dot\varphi {d \over dt}
(\xi \dot\varphi) + {i \over 2} \l
{d \over dt} ( \xi \dot\l) + {i \over 2} \xi \dot\l \dot\l \right] dt.
\eqae
The third term vanishes as $\dot\l \dot\l =0$, and the second term
vanishes after partial integration,
whereas the first term yields after partial integration of one-half
of this term (this ${1 \over 2} - {1
\over 2}$ trick cancels the $\ddot\varphi$ terms)
\eqa
\delta S \; {\rm (rigid)} \; = \int \left[ {1 \over 2} \dot\xi
\dot\varphi \dot\varphi + {d \over dt}
\left( {1 \over 2} \xi \dot\varphi \dot\varphi + {i \over 2} \xi \l
\dot\l \right) \right] dt.
\eqae
We see that $\dot\varphi \dot\varphi$ is the Noether current for
translations for the $(\varphi, \l)$ system.  (Also the other way to construct the Noether current gives the same result).

Introducing the gauge field $h$ for gravity and defining $\delta h = {1 \over 2} \dot\xi + \cdots$ we
obtain
\eqa
L \; {\rm (Noether)} \; =- h \dot\varphi \dot\varphi, \quad \delta h
= {1 \over 2} \dot\xi.
\eqae
Note that there is no term of the form $h \l \dot\l$ in $L$
(Noether).  (However
the field redefinition (\ref{1thirtyeight}) produces a $h \l \dot\l$ term
in the action in (\ref{action})).  As Noether current for $\l$ one might have expected $\pi \dot\l$ where $\pi$ is the conjugate momentum of $\l$, but $\dot\l$ vanishes on-shell, and anyhow time-derivatives of fields should not appear in the Hamiltonian formalism.

Variation of $\dot\varphi$ in $L$ (Noether) yields
\eqa
\delta S \; {\rm (Noether)} \; = \int - 2 h \dot\varphi {d \over dt}
(\xi \dot\varphi) dt,
\eqae
which can be canceled, using the $\tfrac{1}{2} - \tfrac{1}{2}$ trick, by $\delta h =
\xi \dot{h} - h \dot\xi$.  We could also
have proceeded in another way: if we would partially integrate we
would obtain $2 \dot{h} \dot\varphi \xi
\dot\varphi + 2 h \ddot\varphi \xi
\dot\varphi$, and the first variation could be removed by $\delta h =
2 \xi \dot{h}$ while the second
variation could be eliminated by a suitable $\delta \varphi = 2 \xi h
\dot\varphi$ in ${1 \over 2} \dot\varphi^2$.
However, the variation $\delta \varphi =2 \xi h \dot\varphi$ produces
in $S$ (Noether) with the $\tfrac{1}{2} - \tfrac{1}{2}$ trick a new
variation $-2 \dot\xi h \dot\varphi h \dot\varphi$, which can be
canceled in two ways, etc.  All these ambiguities (and more)
are equivalent to field redefinitions.

We have thus shown that
\eqa
L = {1 \over 2} \dot\varphi^2 + {i \over 2} \l \dot\l - h \dot\varphi^2
\label{over}
\eqae
is invariant under
\eqa
\delta \varphi = \xi \dot\varphi , \delta \l = \xi \dot\l , \delta h
= {1 \over 2} \dot\xi + \xi \dot{h}
- \dot\xi h.
\label{varphi}
\eqae
In particular the result for $\delta h$ shows that the $\hat\xi$
dependent terms in (\ref{1thirtyseven}) are
indeed a general coordinate transformation.

In higher dimension the coupling of a scalar field to gravity is given by
\[
L=-{1 \over 2}(-\det g)^{1 \over 2}g^{\mu\nu}\del_\mu \varphi
\del_\nu\varphi,
\]
so in one dimension $e_m^\mu$ corresponds to $1-2 h$.  Then (\ref{varphi}) agrees with the usual transformation rules of general relativity.

Adding the coupling to the gravitino $-i \psi \dot\varphi \l$, we
obtain invariance under $\xi$
transformations, provided we choose $\delta \psi$ appropriately,
\eqa
\delta (-i \psi \dot\varphi \l) =- i \delta \psi \dot\varphi \l-i
\psi \left[ {d \over dt} (\xi \dot\varphi ) \right]
\l -i \psi \dot\varphi \xi \dot\l.
\eqae
Partially integrating the last two terms, we obtain
\eqa
-i \delta \psi \dot\varphi \l + i \dot\psi \xi \dot\varphi \l,
\eqae
and these two terms can be canceled by choosing $\delta \psi = \xi
\dot\psi$.  In general, Lagrangian
densities in general relativity transform as coordinate densities,
which means that
\eqa
\delta (-i \psi \dot\varphi \l) = {d \over dt} (-i \xi \psi \dot\varphi \l).
\eqae
This is indeed achieved if $\psi$ transforms like
\eqa
\delta \psi = \xi \dot\psi.
\label{dot}
\eqae

Thus the model in (\ref{1thirtythree}) has now also been shown to be invariant under
general coordinate transformation. The result in (\ref{dot}) can be explained
by noting that the field $\psi$ in the Noether coupling in (\ref{1already}) corresponds to
$(\det g)^{1 \over 2} g^{\mu\nu} \psi_\nu \sim e_m{}^\mu \psi_\mu$ in one
dimension.  This field $e^\mu_m \psi_\mu$ is a coordinate scalar, in agreement with (\ref{dot}).

Let us again rescale $\l = \sqrt{1+2 hx} \tilde\l$ to obtain an
action with a term $h \tilde\l \dot{\tilde\l}$.  We obtain from (\ref{over})
\eqa
&& L = {1 \over 2} \dot\varphi^2 + {i \over 2} \tilde\l
\dot{\tilde\l} -h \left( \dot\varphi^2 - i x \tilde\l {d
\over dt} \tilde\l \right), \nn
&& \delta \tilde\l = {\delta \l \over \sqrt{1+2 h x}} - {\delta h x
\l \over (1+2 hx)^{3/2}} = \xi {d \over dt} \tilde\l -
{1 \over 2} x \dot\xi \left( {1-2h \over 1+2 hx} \right) \tilde\l.
\eqae
For $x =-1$ we find again a polynomial result $\delta \tilde\l = \xi
\dot{\tilde\l} + {1 \over 2} \dot\xi \tilde\l$. The second term indicates
that $\tilde\l$ is a half-density, in agreement with our earlier
observation that $1-2h=e_m{}^\mu=\det e_m{}^\mu$ is a density, and $\tilde\l = \sqrt{1-2 h} \l$ is a half-density.

We already saw that it is natural to rescale $\psi$ as in
(\ref{1forty}).  After the rescaling $\psi \sqrt{1+2
hx} = \tilde\psi$ according to which $-i \psi \dot\varphi \l = -i
\tilde\psi \dot\varphi \tilde\l$, the field $\tilde\psi$ transforms
for $x =-1$ as
\eqa
\delta \tilde\psi = \xi {d \over dt} \tilde\psi - {1 \over 2} \dot\xi
\tilde\psi.
\eqae

Collecting all results for $x=-1$, and dropping the tildas, we have
found the following action and
transformation rules for general coordinate transformations
\eqa
&& L = {1 \over 2} \dot\varphi^2 + {i \over 2} \l \dot\l -h
(\dot\varphi^2 + i \l \dot\l)- i \psi \dot\varphi \l,
\nn
&& \delta \varphi = \xi \dot\varphi , \quad \delta \l = \xi \dot\l + {1
\over 2} \dot\xi \l,\quad \delta h = {1
\over 2} \dot\xi  + \xi \dot{h}  - \dot\xi h , \quad \delta \psi = \xi
\dot\psi - {1 \over
2} \dot\xi \psi.
\label{general}
\eqae
The susy transformation rules for this model were given in (\ref{delta}) and (\ref{polynomial}),
\eqa
&& \delta \varphi = i \epsilon (1-2 h) \l,\quad \delta \l = -(1-2 h)
\dot\varphi \epsilon,\nn
&& \delta \psi = (1-2h) [ (1-2 h) \dot\epsilon - \dot{h} \epsilon ], \quad
\delta h = -(1-2 h) i \epsilon \psi.
\label{1sixty}
\eqae

Hence we have found two polynomial formulations of this supergravity model,
one without a coupling $h \l \dot\l$ in (\ref{1thirtythree}), and one with it in (\ref{general}).
If one varies
the susy Noether current $\dot\varphi \l$ in flat
space under rigid susy, one finds
\eqa
\delta ( \epsilon ) \dot\varphi \l = {d \over dt} (i \epsilon \l) \l
- \dot\varphi \dot\varphi \epsilon =
-(\dot\varphi \dot\varphi + i \l \dot\l ) \epsilon.
\label{1sixtyone}
\eqae
This is the current which couples to $h$ in the model with a $h \l
\dot\l$ coupling.  The action of this
model can suggestively be written as
\eqa
L = {1 \over 2} (1-2 h) (\dot\varphi^2 +i \l \dot\l ) - i \psi \dot\varphi
\l.
\label{1sixtyfoour}
\eqae
It is this formulation of the model which is easiest to write in superspace.

\mathversion{bold}
\section{Rigid and local $N=1$ superspace}
\mathversion{normal}

We now turn to the superspace description.  
The coordinates of the superspace for our toy model are $t$ and $\theta$; both are real, and $\theta$ is a Grassmann variable. 
 A superfield $\Phi$ depends on $t$ and $\theta$, but a Taylor expansion of $\Phi$ in terms of $\theta$ contains only two terms because $\theta \theta =0$.  We begin with
\eqa
\Phi = \varphi + i \theta \l,
\eqae
where $\varphi$ and $\l$ are arbitrary functions of $t$.
Since $\varphi$ is real, also $\Phi$ is real, and this requires the factor $i$.
Susy transformations in flat space are generated by the hermitian%
\footnote{ To show that $Q$ is hermitian, note that from the anticommutator $\{ {\del \over \delta \theta} , \theta \} =1$ 
if follows that $\del / \del \theta$ is hermitian.  
Likewise, it follows from the commutator $[ {\del \over \del t} , t ] =1$ that $\del / \del t$ is antihermitian.} 
 generator $Q$
\eqa
Q= {\del \over \del \theta} + i \theta {\del \over \del t},
\eqae
because
\eqa
\epsilon Q \Phi &=& \epsilon \left( {\del \over \del \theta} + i
\theta {\del \over \del t} \right) (
\varphi + i \theta \l) = i \epsilon \l + i \theta (-\epsilon \dot\varphi)\nn
&=& \delta \varphi + i \theta \delta \l.
\eqae
The results for $\delta \varphi$ and $\delta \l$ agree with (\ref{1five}).

The susy covariant derivative, by definition, anticommutes
with $Q, \{ Q,D \} =0$, and this determines $D= {\del \over \del \theta} - i \theta
{\del \over \del t}$. The action can be written as
\eqa
S &=& {i \over 2} \int dtd \theta  (\del_t \Phi ) (D \Phi) \nn
&=& {i \over 2} \int dtd \theta (\dot\varphi + i \theta \dot\l ) ( i
\l - i \theta \dot\varphi) = {1
\over 2} \int [ \dot\varphi \dot\varphi + i \l \dot\l ] dt.
\eqae
We used $\int d \theta \theta =1$ and $\int d \theta c =0$ if $c$ is
independent of $\theta$. This definition of integration is called the
Berezin integral \cite{Berezinfa} and it follows from translational invariance in $\theta$.
Namely requiring that $\int d \theta f(\theta) = \int d \theta
f(\theta+c)$, and using that $f(\theta)=a + b \theta$ because $\theta
\theta=0 $, one finds $\int d \theta c =0$. We normalize $\theta$ such
that $\int d \theta \theta = 1$.

The susy invariance of the action follows from $\delta \Phi =[\epsilon
Q, \Phi]$, $\del_t \epsilon Q = \epsilon Q \del_t$ and $D \epsilon Q =
\epsilon Q D$. One gets then for the variation of the action an expression
of the form $\delta S = {i \over 2} \int d t d \theta Q[...]$. The
${\del \over \del\theta}$ in $Q$ yields zero because $\del_\theta [\dots]$
contains no $\theta$ and $\int d \theta c = 0$.  The $i \theta {\del
\over \del t}$ in $Q$ yields zero because of the total t-derivative.  This shows that any action built from superfields and derivatives ${\del \over \del t}$ and $D$ is always supersymmetric.

After performing the $\theta$ integration, we have obtained the correct $t$-space action.  We can also show before the $\theta$ integration that $\partial_t \phi D \phi $ is the correct Lagrangian density by checking that it has the correct dimension:  the dimension of $\phi$ is that of $\varphi$, so $[ \phi ] =- 1/2$, and $[ dt ] =-1$ and $d \theta = 1/2$.  The action should be dimensionless, so we need derivatives acting on $\phi$ with dimension $+3/2$.  The only derivatives we have available are ${\del \over \del t}$ and $D$, so the Lagrangian density is unique.
If one considers generalized unitary group elements $e^{\theta Q + i t H}$, one finds by
left-multiplication the vielbein fields
\eqa
&& e^{\epsilon Q +iaH} e^{\theta Q +itH}= e^{\epsilon Q +iaH +\theta Q +itH
+{1 \over 2} [ \epsilon Q, \theta Q ]} \nn
&& \equiv \exp [ \{ z^\Lambda + dz^M E_M{}^\Lambda (Z) \} ] T_\Lambda.
\label{equiv}
\eqae
We used the Baker-Campbell-Hansdorff formua, and $z^\Lambda = (t, \theta ), T_\Lambda = (H,Q)$ and $dz^M = (a, \e)$.
Using $[\epsilon Q, \theta Q]=- \epsilon \{Q,Q\} \theta= - 2\epsilon H
\theta$ we find in the exponent $(\theta + \epsilon)Q + i (t + a +
i\epsilon \theta)H$.  Thus $\delta (\epsilon) \theta = \epsilon $ and $\delta (\epsilon) t = i
\epsilon \theta$. This is a nonlinear representation of the susy algebra
\eqa
\{Q,Q\}=2H {\bf ,} \;\; [Q,H]=0, \;\; {\rm and } \;\; [H,H]=0.
\eqae
in terms of coordinates. The field representation in (\ref{1five}) is induced by
this coordinate representation, namely $\Phi ' (t',\theta ') = \Phi
(t,\theta)$.%
\footnote{Actually, one finds in this way minus the result of (2.6). To get (2.6), we would have replaced $\epsilon$ (and $a$) in (4.5) by $-\epsilon$ (and $-a$). In general, coordinates transform contragradiently (opposite) to fields.}

One can repeat the same procedure as in (\ref{equiv}) but now using right multiplication.
Right multiplication yields $D$, and now we understand why $\epsilon_1 D$ and $\epsilon_2 Q$ commute: because left and
right multiplications commute.

In 4 dimensions it is better to define $x$-space components by $D$
derivatives of $\Phi$ at $\theta
=0$.  Here this makes no difference since there are only two components in $\Phi$
\eqa
\Phi_{\big| \theta =0} = \varphi (t), \quad D \Phi_{\big| \theta =0} = i \l (t).
\eqae
Since $\int dt \int d \theta$ can be replaced
by $\int dt D$ (recall that $\int d \theta = {\del \over \del \theta}$
and $\int dt i \theta {\del \over
\del t} L =0$ because it is a total $t$-derivative) we get
\eqa
S &=& {i \over 2} \int dt [ (\del_t D \Phi) D \Phi + \del_t \Phi DD \Phi
]\nn
&=& {i \over 2} \int dt [ (i \dot\l ) (i \l) -  \dot\varphi i  \dot\varphi ].
\eqae

The extended susy algebra of $Q, D$ and $H = i {\del \over \del t}$ reads
\eqa
&& \{ Q, Q \} = 2 H,\;  \; \{ D,D \} =- 2 H, \; \; \{ Q, D \} =0, \nn
&& [H, Q] = [ H,D ] =0.
\eqae

To formulate also the supergravity model in superspace, the $x$-space
action of (3.43)
\eqa
L = {1 \over 2} (1-2 h) (\dot\varphi^2 +i \l \dot\l) -i \psi \dot\varphi\l,
\eqae
suggests to introduce as superfield for the gauge fields of supergravity
\eqa
H = ( 1-2 h) +2i \theta \psi,
\label{2seventytwo}
\eqae
and to write
\eqa
L  = { i \over 2} \int d \theta H (\del_t \Phi) D \Phi.
\label{1seventyfour}
\eqae
This is the action in the so-called prepotential approach.
This action is not manifestly covariant w.r.t.\ general coordinate
transformations in superspace, but we
shall soon give another formulation which is manifestly covariant and
equivalent, giving the same
$t$-space action.  Let us first check that the expression in (\ref{1seventyfour}) yields
indeed the $t$-space action
\eqa
L &=& {i \over 2} \int d \theta [ (1-2h) +2i \theta \psi ] (
\dot\varphi + i \theta \dot\l )(i \l - i \theta
\dot\varphi)  \nn
&=& {1 \over 2} (1-2h) (\dot\varphi \dot\varphi + i \l \dot\l ) - i
\psi \dot\varphi \l.
\eqae
This agrees with (\ref{1sixtyfoour}).

We now turn to the covariant approach.  One introduces ``flat
covariant derivatives"
\eqa
\cd_{\bar{t}} &=& E_{\bar{t}}{}^\theta D_\theta + E_{\bar{t}}{}^t  \del_t ,\nn
\cd_{\bar\theta} &=& E_{\bar\theta}{}^\theta D_\theta +
E_{\bar\theta}{}^t  \del_t.
%\cd_\theta &\equiv& \cd.
\eqae
The superfields $E_{\bar{t}}{}^\theta$ etc are the inverse super vielbeins.
The bars on $\bar\theta$ and $\bar{t}$ denote that these are flat
indices, while $\theta$ and $t$ are curved
indices.  Because in rigid susy the natural derivatives are $\del_t$
and $D_\theta$ rather than $\del_t$ and
$\del_\theta$, one introduces vielbeins and parameters on this basis.
We introduce the notation $D_\L = \{ D_\theta ,
\del_t \}$ and $\cd_M = \{ \cd_{\bar\theta}, \cd_{\bar{t}} \}$ where
$\L = \{ \theta, t \}$ and $M = \{
\bar\theta, \bar{t} \}$.  Then the super-vielbeins on the basis
$\del_\Lambda =(\del_\theta , \del_t)$ are related to the vielbeins on the
basis $\cd_\Lambda$ as follows
\eqa
&& \tilde{E}_M{}^\L  \del_\L = E_M{}^\L D_\L,\;\; \tilde{E}_M{}^\theta =
E_M{}^\theta, \;\; {\rm but} \;\;
\tilde{E}_M{}^t = E_M{}^t - E_M{}^\theta i \theta, \nn
&& \tilde\Xi^\L \del_\L = \Xi^\L D_\L,\;\; \tilde\Xi^\theta = \Xi^\theta,
\;\; {\rm but}\; \; \tilde\Xi^t =
\Xi^t -  \Xi^\theta i \theta.
\label{1seventyseven}
\eqae

A supercoordinate transformation of $\Phi$ becomes $\delta \Phi =
\tilde\Xi^\L \del_\L \Phi = \Xi^\L D_\L
\Phi$.  The (inverse) super vielbein $\tilde{E}_M{}^\L$ transforms as
in ordinary general relativity.
\eqa
\delta \tilde{E}_M{}^\L &=& \tilde\Xi^\Pi (\del_\Pi \tilde{E}_M{}^\L) -
\tilde{E}_M{}^\Pi (\del_\Pi
\tilde\Xi^\L) \nn
&=& \Xi^\Pi (D_\Pi \tilde{E}_M {}^\L) - E_M{}^\Pi (D_\Pi \tilde\Xi^\L).
\eqae
Contracting with $\del_\L$ on both sides of the equation and then
using (\ref{1seventyseven}) one obtains
\eqa
\delta E_M{}^\S = \Xi^\Pi D_\Pi E_M{}^\S  + E_M{}^\Pi D_\Pi \Xi^\S +
E_M{}^\Pi \Xi^\L T^{(0)}_{\L \Pi}{}^\S,
\eqae
where $[D_\L , D_\Pi \} = T^{(0)}_{\L \Pi}{}^\S  D_\S$.  In our
case only the $\theta\theta$ component of the supertorsion is
nonvanishing, $T^{(0)\; t}_{\theta\theta} = -2i$ (see (4.9)), and we get
\eqa
\delta E_M{}^t &=& \Xi^\Pi D_\Pi E_M{}^t - E_M{}^\Pi (D_\Pi \Xi^t +
2i \Xi^\theta \delta_\Pi{}^\theta), \nn
\delta E_M{}^\theta &=& \Xi^\Pi D_\Pi E_M{}^\theta - E_M{}^\Pi D_\Pi \Xi^\theta.
\eqae

In $x$-space we only have the field content of $H$. To reduce the extra
superfields we shall now impose a constraint on the super-vielbein and
choose a gauge. Afterward we shall come back to the general super
coordinate transformations in the presence of this constraint and gauge.

To avoid cumbersome notation we write $\cd$ for $\cd_{\bar\theta}$ and
parametrize $\cd$ as follows
\eqa
\cd = E D + X i \del_t, \quad {\rm with} \quad D=D_{\theta} = {\del \over \del \theta}
- i \theta {\del \over \del t}.
\eqae
Thus $E_{\bar\theta}{}^\theta$ is denoted by $E$, and
$E_{\bar\theta}{}^t$ by $i X$.  The fields in
$\cd_{\bar{t}}$ are determined in terms of the fields in $\cd$ by imposing
the following \emph{constraint}.
\eqa
{1 \over 2} \{ \cd , \cd \} = -i \cd_{\bar{t}}.
\label{1eightyone}
\eqae
This is the curved space extension of $\{ D, D \} =- 2 i \del_t$.  It
is a so-called conventional constraint; in 4 dimensions
it expresses the bosonic Lorentz connection in terms of the fermionic
Lorentz connections, but here no connection is present,
and the constraint expresses the super vielbeins with flat index
$\bar{t}$ in terms of those with flat index $\bar\theta$.

The anticommutator in (\ref{1eightyone}) yields
\begin{gather}
{1 \over 2} \{ ED + X i \del_t, ED + Xi \del_t \} \nonumber \\
= [-E^2 + E (D X)
+ i X \dot{X} ] i \del_t + [E (DE) + i X \dot{E} ]D \nonumber \\
\equiv -i [E_{\bar{t}}{}^{\theta} D + E_{\bar{t}}{}^{t} \del_{t}].
\end{gather}
Hence
\eqa
E_{\bar{t}}{}^t &=& E^2 - E (DX) - i X \dot{X}, \nn
E_{\bar{t}}{}^\theta &=& i E (DE) - X \dot{E}.
\eqae
As a matrix, the inverse super vielbein is thus given by
\eqa
E_M{}^\L = \left( \begin{array}{ccc} E^2 - E (DX) - i X \dot{X}
&\qquad  & i E (DE) -  X \dot{E} \\ i X & & E
\end{array} \right).
\eqae
The superdeterminant of $\left( \begin{array}{cc}  A & B \\ C & D
\end{array} \right)$ being given by%
\footnote{ Use $\left(
\begin{array}{cc}  A & B \\ C & D \end{array} \right) = \left(
\begin{array}{cc}  A-BD^{-1} C
& BD^{-1} \\ 0 & 1 \end{array} \right) \left( \begin{array}{cc}  1 &
0 \\ C & D \end{array} \right)$ and take the product of
the \\two superdeterminants.}
\eqa
s \det \left( \begin{array}{cc}  A & B \\ C & D \end{array} \right) =
{1 \over D} (A - B D^{-1} C),
\eqae
and we find
\eqa
s \det E_M{}^\L &=& {1 \over E} \left\{ E^2 - E (DX)  - i X \dot{X} +
[E (DE) + i X \dot{E} ] {1 \over E} X
\right\}
\nn &=& E - (DX) - i {X \dot{X} \over E} + {DE \over E} X,
\eqae
where we used that $XX=0$.

We now choose \emph{the gauge which sets $i X = E_{\bar\theta}{}^t$ to zero}.  This will fix some
of the gauge symmetries in superspace as we discuss
later.  In the covariant formalism, an invariant action (an action
invariant under super general coordinate
transformation, there are no Lorentz transformations) is given by ($s
\det E_\L{}^M$) times covariant
derivatives on tensors (scalars in our case) with flat indices.  We
find then for the covariant action in
superspace, which generalizes (4.12),
\eqa
S &=&  {i \over 2} \int dt d \theta ( s \det E_\L{}^M )
(\cd_{\bar{t}} \Phi ) (\cd_{\bar{\theta}} \Phi ) \nn
&=& \int dt d \theta \left(  {i \over 2E} \right) (E^2 \del_t \Phi )
(ED \Phi ).
\eqae
(A term $i E(DE) (D\Phi)$ in $\cd_{\bar{t}} \Phi$ cancels out).
Identifying $E^2 =H$ we find the same action as in (\ref{1seventyfour}).

Let us now discuss the local symmetries in superspace.  On $\Phi$ a
general supercoordinate transformation would read
\eqa
\delta \Phi = \tilde\Xi^t \del_t \Phi + \tilde\Xi^\theta \del_\theta \Phi.
\eqae
To incorporate rigid susy as the flat superspace limit $E=1, X=0$, we
expand instead in terms of $\del_t$ and $D_\theta$.
Then
\eqa
\delta \Phi = \Xi^t \del_t \Phi + \Xi^\theta D_\theta \Phi = \Xi \Phi\;
\; {\rm with} \; \;\Xi = \Xi^t \del_t + \Xi^\theta D_\theta.
\eqae
We used again (4.15).  But $\Xi^\theta$ is not a free super parameter as we now show.  We can also write $\Xi \Phi$ as the operator equation $[ \Xi, \Phi ] = 0$.

The transformation rules of the super vielbein follow from
\eqa
\delta \cd_M = [ \Xi , \cd_M ].
\eqae
In particular for $\cd_{\bar{\theta}} = \cd = ED + X i \del_t$ we find
\eqa
\delta [ED +X i \del_t ] = [ \Xi^t \del_t + \Xi^\theta D, E D + X i \del_t].
\eqae
If the gauge $X=0$ has been reached, the terms with $\del_t$ on the
right-hand side should vanish because they are absent on the left-hand
side. This yields
the following relation between $\Xi^\theta$ and $\Xi^t$
\eqa
-\Xi^\theta E 2i \del_t - E (D \Xi^t ) \del_t =0 \,\Rightarrow\,
\Xi^\theta = {i \over 2} (D \Xi^t).
\eqae
Expanding
\eqa
\Xi^t = \xi (t) -2 i \theta \epsilon (t),
\eqae
we find
\eqa
\Xi^\theta = {i \over 2} (-2i \epsilon (t) - i \theta \dot\xi ) =
\epsilon (t) + {1 \over 2} \theta \dot\xi.
\eqae

The transformation of $\Phi$ becomes then
\eqa
\delta \Phi = (\xi - 2 i \theta \epsilon ) \dot\Phi + \left( \epsilon
+ {1 \over 2} \theta \dot\xi \right)
  D \Phi,
\eqae
which reads in components
\eqa
\delta \varphi &=& \xi \dot\varphi + i \epsilon \l, \nn
\delta \l &=& \xi \dot\l + {1 \over 2} \dot\xi \l - 2 \epsilon
\dot\varphi +  \epsilon \dot\varphi.
\eqae
These are the correct $x$-space results of (3.38) and (\ref{1sixty})
 after defining $\epsilon = \epsilon (x- {\rm space}) (1 -2 h)$.

To obtain the transformation rules of the supergravity gauge fields
in $E$, we return to
\eqa
\delta \cd_M = [ \Xi, \cd_M ] \; {\rm with} \; \Xi = \Xi^t \del_t +
\Xi^\theta D.
\eqae
From this relation $\delta E$ follows if one takes $\cd_{\bar{\theta}} =
\cd = ED$
and collects all terms with $D$,
\eqa
( \delta E) D &=& ( \Xi E) D -E (D \Xi^\theta )D, \nn
\delta E &=& \Xi^t \dot{E} + \Xi^\theta (DE) - E(D \Xi^\theta),\nn
\delta H &=& 2 E \delta E = \Xi^t \dot{H} + \Xi^\theta DH - 2H (D \Xi^\theta).
\label{2ninety7}
\eqae
Substituting the component expressions of (\ref{2seventytwo}) leads to
\eqa
&& - 2 \delta h + 2i \theta \delta \psi = ( \xi -2 i \theta \epsilon)
(-2 \dot{h} +2i  \theta \dot\psi ) + \left(
\epsilon + {1
\over 2} \theta \dot\xi \right) (2 i \psi +2i \theta \dot{h} )\nn
&& \quad - 2 [ (1-2 h) +2i \theta \psi ] \left( {1 \over 2} \dot\xi -
i  \theta \dot\epsilon \right).
\eqae
This leads to
\eqa
\delta (-2h) &=& -2 \xi \dot{h}  +2i \epsilon \psi - (1-2 h) \dot\xi, \nn
\delta (2 \psi) &=&  4 \epsilon \dot{h} +2 \xi \dot\psi + \dot\xi
\psi - 2 \dot{h} \epsilon \nn
&& - 2 \dot\xi \psi +2 (1-2 h) \dot\epsilon.
\label{doot}
\eqae
These results agree with (3.41) using again the identification $\epsilon =
\epsilon (x- \; {\rm space} \; ) (1-2 h)$.

Instead of choosing a gauge we can also find a redefinition of the
fields which leads to the same result.  We could demonstrate
this in the most general case, but to simplify the analysis we still
impose the constraint ${1 \over 2} \{ \cd , \cd \} = -i
\cd_{\bar{t}}$.  Hence $E_{\bar{t}}{}^t$ and $E_{\bar{t}}{}^\theta$
are given in terms of $E$ and $X$.  The action in the gauge
$X=0$ is given by
\eqa
L= \int d \theta \left(  {i \over 2} \right) E^2 (\del_t \Phi ) (D \Phi ).
\label{2one01}
\eqae
whereas the action without this gauge choice reads
\begin{gather} \label{2one02}
L = {i \over 2} \int d \theta \left( E-DX - i X \dot{X} /E + {DE
\over E} X \right)^{-1} \nonumber \\
[ (E^2 - E(DX) -i X \dot{X} ) \dot\Phi + ( i EDE -  X
\dot{E} ) D \Phi ) ] [ ED \Phi + i X \dot\Phi ].
\end{gather}
The question we ask is this: which field redefinition (redefinition
of $E, X, \Phi$) leads from (4.41) to (\ref{2one01})?  The
solution of this question is most easily found by first making a
general symmetry transformation on the fields in (4.41) and
then requiring that the new field $X'$ vanishes.  This will express
the symmetry parameter in terms of $E$ and $X$, and provide an
explicit expression for the new field $E'$ in terms of $E$ such that
the action takes the form in (\ref{2one01}).

The symmetries of the theory are in this case general supercoordinate
transformations.  In classical general relativity a contravariant supervector $F^\L$
transforms, by definition, as follows:
$(F^\L )' (Z') = F^\Pi (Z) {\del \over \del Z^\Pi} (Z')^\L$. Since
field redefinitions do not change $Z$ into $Z'$, we
need a way to write general supercoordinate transformations as
relations between $(F^\L)' (Z)$ and $F^\L (Z)$ \emph{at the
same point $Z$}.  Such a formalism exists and we now explain it and
then will use it. One may call it the ``operator approach to diffeomorphisms''.

For a scalar superfield (a scalar with respect to general
supercoordinate transformations) we define, as it is customary for internal symmetries in 
Yang-Mills theory,
\eqa
\phi' (Z) = e^{\hat{X} (Z) D} \phi (Z).
\label{2one03}
\eqae
Here $Z= \{ \theta, t\}$ and $\hat{X} (Z)$ is an arbitrary superfield
parameter which we will later write as a complicated
expression in terms of $E$ and $X$.  The most general transformation would
be
\eqa
\phi' (Z) = e^{\hat{X} D + \hat{Y} \del_t} \phi (Z).
\eqae
but we will reach our aim with $\hat{Y} =0$.  There will then still
remain super-diffeomorphisms with one superfield parameter
which keep $X' =0$.  They correspond to $\epsilon$ and $\xi$
transformations in $x$-space.

Usually one writes a general supercoordinate transformation as
\eqa
\phi' (Z') = \phi (Z),\quad Z' = Z' (Z).
\label{2one05}
\eqae
The relation between (\ref{2one05}) and (\ref{2one03}) becomes clear
if one writes the latter as
\eqa
e^{-\hat{X} (Z) D} \phi' (Z) = \phi' (Z').
\label{2one06}
\eqae
A particular case is $e^{- \hat{X} (Z)D} Z= Z'$. This yields the relation between 
$Z'(Z)$ and $\hat{X}(Z)$. To work this out we expand
\eqa
&& e^{\hat{X} (Z) D} = 1+ \hat{X} D + {1 \over 2} \hat{X} D \hat{X} D
+ {1 \over 3!} \hat{X} D \hat{X} D \hat{X} D \nn
&& = 1 + \hat{X} D + {1 \over 2} \hat{X} (D \hat{X} ) D + {1 \over
3!} \hat{X} (D \hat{X}) (D \hat{X} ) D + \cdots \nn
&& = 1 + \hat{X} \left( {e^{(D \hat{X})} -1 \over (D \hat{X})}
\right) D \equiv 1 + f \hat{X} D,
\eqae
where $f = f ((D \hat{X})) = (e^{(D \hat{X})} -1) / (D \hat{X})$.  We
used that $\hat{X} \hat{X} =0$, because $\hat{X}$ is anticommuting.  It follows that
\eqa
Z' = e^{- \hat{X} D} Z= (1- \bar{f} \hat{X} D) Z=Z- \bar{f} \hat{X} (DZ),
\eqae
where $\bar{f} \equiv f  (-(D \hat{X})) = 1- {1 \over 2} (D \hat{X} )
+ \cdots$.  More explicitly,
\eqa
\theta' = \theta - \bar{f} \hat{X} (D \theta) = \theta - \bar{f}
\hat{X},\quad t' =t - \bar{f} \hat{X} (D t) = t + \bar{f} \hat{X} (i
\theta).
\eqae
Then we find indeed (\ref{2one06})
\eqa
 e^{- \hat{X} D} \phi' (Z) &=& (1- \bar{f} \hat{X}D ) \phi ' = \phi '
- \bar{f} \hat{X} \left( {\del \over \del \theta} - i
\theta {\del \over \del t} \right) \phi ' \nn
& = &\phi' ( \theta - \bar{f} \hat{X} , t +\bar{f} \hat{X} i \theta )
= \phi' (Z').
\eqae

Consider now a transformation with $\hat{X}$ which is such that the
new field $X'$ vanishes.  In terms of
\eqa
\cd = ED + iX \del_t,
\eqae
this means that
\eqa
\cd' = e^{\hat{X} D} \cd e^{- \hat{X} D} = E' (Z) D.
\eqae
Thus in $D'$ all terms with free derivatives $i \del_t$ and $i
\del_t D$ must cancel, and the coefficient of $D$ is by
definition $E' (Z)$.  Requiring that the coefficients of $i \del_t$
and $i \del_t D$ vanish fixes $\hat{X}$ as a function of
$X$ and $E$.

The details are as follows.  We begin with
\eqa
e^{\hat{X} D} \cd e^{- \hat{X} D} &=& (1 + f \hat{X} D) (ED +X i
\del_t)(1- \bar{f} \hat{X} D)\nn
&=& E' (Z) D + X' (Z) i \del_t.
\label{2one12}
\eqae
Somewhat laborious algebra using $DD = - i \del_t$ yields the following
result for this expression,
\eqa
&& [ \{ E+ f \hat{X} (DE) \} D+ \{ X - f \hat{X} E+ f \hat{X} (DX) \}
i \del_t \nn
&& - f \hat{X} X i \del_t D] (1- \bar{f} \hat{X} D) \nn
&& = [ E + f \hat{X} (DE - ED ( \bar{f} \hat{X} ) - f \hat{X} (DE)
\bar{f} (D \hat{X}) \nn
&& - X i \del_t (\bar{f} \hat{X} ) - f \hat{X} E \bar{f} (i \del_t
\hat{X} ) - f \hat{X} (DX) \bar{f} (i \del_t \hat{X} ) ] D
\nn
&& + [ X - f \hat{X} E + f \hat{X} (DX) - E \bar{f} \hat{X} ] i \del _t \nn
&& + [ - f \hat{X} X - X \bar{f} \hat{X} +f\hat{X}X\bar{f}(D\hat{X})] D i \del_t.
\eqae
The terms with $Di\del_t$ should vanish identically, and one can check that this is the
case. The vanishing of the terms with $i \del_t$ shows that $X$ is
proportional to $\hat{X}$, and then the terms with $D i \del_t$
also vanish.  Hence
\eqa
&& X = f \hat{X} E - f \hat{X} (DX) + E \bar{f} \hat{X}, \nn
&& (1 + f \hat{X} D) X =(f + \bar{f}) E \hat{X} ,\nn
&& e^{\hat{X} D} X = (f + \bar{f}) E \hat{X}, \nn
&& X = e^{- \hat{X} D} [(f + \bar{f} ) \hat{X} E].
\eqae
This relation expresses $X$ in terms of $\hat{X}$ and $E$, but we
need $\hat{X}$ in terms of $X$ and $E$.  We can invert by
expanding the right-hand side and then solve for $\hat{X}$ iteratively
\eqa
X &=& - (f + \bar{f} ) \hat{X} E + \hat{X} D [(f + \bar{f} ) \hat{X}
E ] + \cdots \nn
&=& - \left(2 + {1 \over 3} (D \hat{X})^2 + {1 \over 60} (D
\hat{X})^4 + \cdots \right) \hat{X} E \nn
&& + \hat{X} D \left[ 2 \hat{X} E + {1 \over 3} (D \hat{X} )^2
\hat{X} E + {1 \over 60} (D \hat{X} )^4 \hat{X}
E + \cdots \right] \nn
&=& -2 \hat{X} E + 2 \hat{X} (D \hat{X}) E \; .
\eqae
Inversion of $\hat{X} - \hat{X} (X \hat{X}) = - {1 \over 2} X/E
\equiv - \tilde{X}$ yields
\eqa
&& \hat{X} = - \tilde{X} + \tilde{X} (D \tilde{X} ) - 2 \tilde{X} (D
\tilde{X})^2 \nn
&& \quad 5 \tilde{X} (D \tilde{X} )^3 + 14 \tilde{X} (D \tilde{X})^4
+ \cdots \quad {\rm with} \quad \tilde{X} \equiv {1 \over 2} {X \over E}.
\eqae

This expression for $\hat{X}$ can then be used to find the new field $E'$
\eqa
E' &=& E + f \hat{X} (DE) - E D (\bar{f} \hat{X} ) + \cdots \nn
&=& E- {X \over E} (DE) + E D ({X \over E} ).
\label{2one17}
\eqae
If one then substitutes this $E'$ and the $\Phi'$ in (4.42) into (4.40), the result is
equal to (4.41).

It may be helpful for becoming more familiar with the operator
approach to diffeomorphisms to check that one gets the same
results as from the usual approach.  In the usual approach one begins
with supervielbeins (supervectors in supercoordinate space)
$\tilde{E}_M{}^\L$ and general coordinate transformations $Z^\L
\rightarrow Z^\L + \Xi^\L$ where we take $\Xi^\L$ as infinitesimal.
Then
\eqa
\delta \tilde{E}_M{}^\L &=& \tilde\Xi^\Pi \del_\Pi \tilde{E}_M{}^\L -
\tilde{E}_M{}^\Pi \del_\Pi \tilde\Xi^\L \nn
&=& \Xi^\Pi D_\Pi \tilde{E}_M{}^\L - E_M{}^\Pi D_\Pi \tilde\Xi^\L,
\eqae
with $\del_\Pi = \left( {\del \over \del \theta} , {\del \over \del
t} \right)$ and $D_\Pi = \left( \tfrac{\del}{\del \theta} - i
\theta {\del \over  \del t} ,{\del \over \del t} \right)$.  From (4.15)
\eqa
&& \tilde{E}_M{}^\theta = E_M{}^\theta,\quad \tilde{E}_M{}^t = E_M{}^t -
E_M{}^\theta i \theta\nn
&& \tilde\Xi^\theta = \Xi^\theta,\quad \tilde\Xi^t = \Xi^t -\Xi^\theta i \theta.
\eqae
One finds by straightforward substitution
\eqa
\delta E_M{}^\theta &=& \delta \tilde{E}_M{}^\theta = \Xi^\Pi D_\Pi
\tilde{E}_M{}^\theta - E_M{}^\Pi D_\Pi \tilde\Xi^\theta \nn
&=& \Xi^\Pi D_\Pi E_M{}^\theta - E_M{}^\Pi D_\Pi \Xi^\theta.
\eqae
This agrees with (4.36).
However, in $\delta E_M{}^t$ there are extra terms beyond those which
are present in the corresponding relation for $\tilde{E}_M{}^t$.
\eqa
&& \delta E_M{}^t = \delta ( \tilde{E}_M{}^t + E_M{}^\theta i \theta
) = \Xi^\Pi D_\Pi \tilde{E}_M{}^t -E_M{}^\Pi D_\Pi (
\Xi^t - \Xi^\theta i \theta) \nn
&& + ( \Xi^\Pi D_\Pi E_M{}^\theta - E_M{}^\Pi  D_\Pi \Xi^\theta ) i
\theta. %- E_M{}^\theta i \Xi^\theta
\eqae
%where we used that in the last term that $\delta \theta =
%\Xi^\theta$.  
Various terms cancel and one obtains
\eqa
\delta E_M{}^t = \Xi^\Pi D_\Pi E_M{}^t - E_M{}^\Pi D_\Pi \Xi^t - 2 i
E_M{}^\theta \Xi^\theta.
\eqae
The last term is due to the rigid torsion $T_{\theta \theta}{}^t =
-2i$.  Expanding the finite result for $E'$ and $X'$ in
(\ref{2one12}) to first order in $\hat{X} = \Xi^\theta$ we find agreement.

One may also check that
\eqa
\delta \cd &=& ( e^{\hat{X} D} E_{\bar{\theta}}{}^\L e^{-\hat{X} D} ) D_\L +
E_{\bar{\theta}}{}^\L (e^{\hat{X} D} D_\L e^{- \hat{X} D} )
\nn
&=& \delta E_{\bar{\theta}}{}^\L D_\L,
\eqae
but note that $e^{\hat{X} D} E_{\bar{\theta}}{}^\L e^{- \tilde{X} D}$ only
yields the transport terms $E_{\bar{\theta}}{}^\L (Z')$, while the
terms with $e^{\hat{X} D} D_\L e^{-\hat{X} D}$ yield the terms which
are due to the index $\L$ of $E_{\bar{\theta}}{}^\L$.

\section{Extended rigid supersymmetries}

We can also construct models with extended $(N>1)$ susy.   We restrict ourselves to rigid supersymmetry, but $N>1$ supergravities can also be constructed.  We perform the analysis in $x$-space but one could also use $N>1$ superspace.

Consider the action
\eqa
L = {1 \over 2} \dot\phi \dot\phi + {i \over 2} \sum^N_{j=1} \l^j \dot\l^j.
\eqae
It is invariant under the following rigid susy transformations
\eqa
\delta \phi = i \sum^N_{j=1} \epsilon^j \l^j , \delta \l^j = -\dot\phi
\epsilon^j.
\eqae
The proof is as before: $\delta L = \dot\phi {d \over dt} (i
\epsilon^j \l^j) + i \l^j {d \over dt} (\dot\phi \epsilon^j)$
vanishes after partial integration.

There is something unusal about this action: it has more fermions ($N$)
than bosons (one).  Nevertheless it has the same number
of fermionic and bosonic states because after quantization there are
only zero modes $(\hat{x}_0$ and
$\hat{\l}^j_0)$ and their conjugate momenta ($\hat{p}$ and again
$\hat{\l}^j_0$).  The $\hat\l^j_0$ can be combined into $[{N \over 2}]$
annihilation and creation operators $(\hat\l^1_0 \pm i
\hat\l^2_0) / \sqrt{2}$, ... If $N$ is odd, the last $\hat\l^N_0$
yields projection operators $P_{\pm}={1 \over 2}(1 \pm \hat\l^N_0)$ and
the Hilbert space ${\mathcal H}$ split into two spaces ${\mathcal H}=P_{\pm}
{\mathcal H}$. No operator can bring one from a state in ${\mathcal
 H}_+$ to a
state in ${\mathcal H}_-$ because operators either have no $\hat\l^N_0$, or if
they have one $\hat\l^N_0$ then still $\hat\l^N_0 {\mathcal H}_+ ={\mathcal H}_+$
because $\hat\l^N_0 {1 \over 2}(1+ \hat\l^N_0)= {1 \over 2} (1+
\hat\l_0^N)$.
The bosonic states are $e^{ip\hat{x}_0} \mid 0 \rangle = \mid
p \rangle$ but also $\hat\l^1_0 \hat\l^2_0 \mid p \rangle$ etc., and
the fermionic states are $\hat\l^j_0 \mid p \rangle$ but also
$\hat\l^1_0 \hat\l^2_0 \hat\l^3_0 \mid p \rangle$ etc.  In fact there
are $2^{N-1}$ bosonic states (with an even number of
$\hat\l^j_0$) and $2^{N-1}$ fermionic states (with an odd number of
$\l_0^j$).  In higher dimensions there are also oscillators,
and then one can count the number of bosonic and fermionic states by
looking at the number of propagating fields.%
\footnote{ In $d=(1,1)$ one can divide the real scalar into a left-moving piece
 $\varphi (x+t)$ and a right-moving piece $\varphi (x-t)$, 
and susy requires then a two-component spinor $\l = {\l_+\choose\l_-}$ with $\l_+ (x+y)$ right-moving
on-shell and $\l_- (x-t)$ left-moving.  In the zero mode sector there
are always the same number of bosonic and
fermionic states but in the nonzero mode sector one may drop $\l_+$
(or $\l_-$) and still have susy.  Then $\l_-$ transforms
into $\varphi$ and vice-versa, while $\l_+$ is inert.}  For
example, in $d=(3,1)$ one has the Wess-Zumino model with one Majorana
spinor $\l^\a$ (2 states) and two real propagating massless
scalars (again 2 states per oscillator).

In our model similar things happen.  To close the algebra off-shell
we need an equal number of bosonic and fermionic field
components.  Thus for $N=2$ we need one real auxiliary field, which
we call $F$.  The action reads
\eqa
L = {1 \over 2} \dot\varphi \dot\varphi - {i \over 2} (\l^1 \dot\l^1
+ \l^2 \dot\l^2) + {1 \over 2} F^2.
\eqae
One can view $(\varphi, \l^1)$ as one multiplet, and $(\l^2, F)$ as
another.  The action for these multiplets reads in
superspace
\eqa
L (\varphi, \l^1) &=& {1 \over 2} \dot\varphi \dot\varphi - {i \over
2} \l^1 \dot\l^1 \sim \int d \theta D \Phi \dot\Phi,\quad \Phi =
\varphi + i \theta \l^1, \nn
L (\l^2, F) &=& - {i \over 2} \l^2 \dot\l^2 + {1 \over 2} F^2 \sim
\int d \theta \Psi D \Psi, \quad \Psi = \l^2 + \theta F.
\eqae
Clearly $\delta \varphi = i \epsilon^1 \l^1$ and $\delta \l^1 =
\dot\varphi \epsilon^1$ form a closed susy algebra, $\{ Q, Q \}
\sim P$.  Also $\delta \l^2 = F \epsilon^2$ and $\delta F = i
\epsilon^2 \dot\l^2$ forms the same closed algebra.  However, in
$x$-space we can write down more general rules
\eqa
 \delta \varphi &=& \sum^2_{j=1} i \epsilon ^j \l^j , \delta \l^j =
\dot\varphi \epsilon^j + \a^{jk} \epsilon^k F, \nn
\delta F &=& i \epsilon^j \beta^{jk} \dot\l^k,
\eqae
where $\a$ and $\beta$ are arbitrary real matrices.  Invariance of
the action requires
\eqa
\a^T = \beta.
\eqae
The commutator algebra on $\varphi$ yields
\eqa
&& [ \delta (\epsilon_1), \delta (\epsilon_2) ] \varphi = \sum i
\epsilon^j_2 (\dot\varphi \epsilon_1^j + \a^{jk} \epsilon_1^k
F) - (1 \leftrightarrow 2) \nn
&& = (2 i \epsilon^j_2 \epsilon^j_1 ) \dot\varphi + [ i \epsilon_2
(\a + \a^T) \epsilon_1 ] F.
\eqae
Clearly, for $\a$ antisymmetric (hence $\a^{jk} \sim \epsilon^{jk}$)
we find only a translation of $\varphi$.  For $F$ one finds
\eqa
&& [ \delta (\epsilon_1), \delta (\epsilon_2) ] F = i \epsilon^j_2
\beta^{jk} {d \over dt} (\dot\varphi \epsilon^k_1 + \a^{kl}
\epsilon^l_1 F) - (1 \leftrightarrow 2) \nn
&& = [ i \epsilon_2 ( \a+ \a^T) \epsilon_1 ] \ddot{\varphi} + [2i
\epsilon_2 \a^T \a \epsilon_1 ] \dot{F}.
\eqae
We now recognize various symmetries.
\medskip

\noindent (i) $\delta \varphi = z F$ and $\delta F = z
\ddot{\varphi}$ with $z= i \epsilon_2 (\a + \a^T) \epsilon_1$.  This
is an
equation of motion symmetry.  (These are symmetries of pairs of
fields $f$ and $g$ of the form $\delta f = A {\delta S \over \del
g}$ and $\delta g= - A {\delta S \over \delta f}$.  Obviously they
leave the action invariant).\\
\noindent (ii)  The usual translation. \\
\noindent (iii) An extra symmetry $\delta F = \s \dot{F}$ where $\s$
is $2 i \epsilon_2 \a^T \a \epsilon_1 - 2 i \epsilon_2
\epsilon_1$.  This is clearly a symmetry of $L = {1 \over 2} F^2$.

For the fermions the susy commutator yields
\eqa
&& [ \delta (\epsilon_1), \delta (\epsilon_2) ] \l^j = \delta
(\epsilon_1) [ \dot\varphi \epsilon^j_2 + \a^{jk} \epsilon^k_2
F] - (1 \leftrightarrow 2) \nn
&& = - i (\epsilon^l_1 \epsilon^j_2 - \epsilon^l_2 \epsilon^j_1 )
\dot\l^l + [ (\a^{jk} \epsilon^k_2 ) ( i \epsilon^l_1
\beta^{lm} \dot\l^m ) - (1 \leftrightarrow 2) ] \nn
&& = [2 i \epsilon_2^l \epsilon_1^l) \dot\l^j + i (\epsilon^l_2
\epsilon^j_1 \dot\l^l - \epsilon^l_2 \epsilon^l_1 \dot\l^j - 
(1 \leftrightarrow 2) ) + [ \cdots ].
\eqae
(iv) $\delta \l^j = s^{jl} \dot\l^l$ with symmetric and real
$s^{jl}$.  This is again an equation of motion symmetry.

The extra symmetries do not form a closed algebra. Their commutators
generate new equation of motion symmetries, etc.,
etc.  However, the minimal extension of the susy rules with $F$ forms
a closed algebra
\eqa
&& \delta \varphi = i (\epsilon^1 \l^1 + \epsilon^2 \l^2 ) ,\quad \delta
\l^1 = \dot\varphi \epsilon^1 - F \epsilon^2 , \quad\delta \l^2
= \dot\varphi \epsilon^2 + F \epsilon^1, \nn
&& \delta F = i \epsilon^1 \dot\l^2 - i \epsilon^2 \dot\l^1.
\eqae

We can try to construct and $N=2$ superspace by introducing
\eqa
&& \theta = {\theta^1 + i \theta^2 \over \sqrt{2}},\quad \bar\theta =
{\theta^1 - i \theta^2 \over \sqrt{2}} , \quad\bar{D} = {\del \over \del
\theta} +
i \bar\theta \del_t, \quad D = {\del \over \del \bar\theta} + i \theta \del_t, \nn
&& \bar{Q} = {\del \over \del \theta} - i \bar\theta \del_t, \quad Q =
{\del \over \del \bar\theta} -i \theta \del_t , \quad \l = {\l^1 +
i \l^2 \over \sqrt{2}}, \quad \bar\l = {\l^1 - i \l^2 \over \sqrt{2}} ,\nn
&& L = {1 \over 2} \dot\varphi \dot\varphi - i \bar\l \dot\l + {1
\over 2} F^2 ,\quad \Phi = \varphi + i \bar\theta \l + i \theta \bar\l + F \bar\theta \theta, \nn
&& L = - \int d \theta d \bar\theta \bar{D} \Phi D \Phi = - \int d
\theta d \bar\theta [ i \bar\l - F
\bar\theta + i
\bar\theta \dot\varphi - \bar\theta \theta \dot{\bar\l} ] [ i \l + F \theta + i \theta \dot\varphi - \theta \bar\theta
\dot\l ] \nn
&& \quad = -i \bar\l \dot\l + i \dot{\bar\l} \l +
\dot\varphi \dot\varphi + F^2.
\eqae
The $t$-space components of $\Phi$ are given by $\Phi_\mid = \varphi,
D \Phi_\mid = i \l, \bar{D} \Phi_\mid = i \bar\l$ and ${1 \over 2} [
\bar{D}, D] =F$.  The susy rules are generated by $(\bar\epsilon D +
\epsilon \bar{D})$ acting on these components, using the $D$ algebra.

The $N=2$ model has a mass term and a Yukawa coupling
\eqa
&& L = \left[ {1 \over 2} \dot\varphi^2 - {i \over 2} \l_1 \dot\l_1 -
{i \over 2} \l_2 \dot\l_2 + {1 \over 2} F^2 \right] \nn
&& - m ( F \varphi + i \l_1 \l_2 ) + g \left( {1 \over 2} F \varphi^2
+ i \l_1 \l_2 \varphi \right).
\eqae
The corresponding action in $N=1$ superspace is
\eqa
&& S^{(0)} = \left( {-i \over 2} \right) \int dt d \theta [ (D_\theta
\phi ) \del_t \phi + i \psi D_\theta \psi ], \nn
&& \phi = \varphi + i \theta \l_1,\quad \psi = \l_2 + \theta F,
\eqae
\eqa
S (m) &=& - \int dt d \theta m \phi \psi, \nn
S (g) &=& \int dt d \theta {1 \over 2} g \phi \phi \psi.
\eqae
One can write more generally a superpotential term as
\eqa
 S (W) = \int dt d \theta W (\phi) \psi = \int dt (WF + i \l_1 W' \l_2 ),
\eqae
where $W (\phi) = - m \phi + {1 \over 2} g \phi^2 + \cdots$.

One can also write down nonlinear sigma models for the $N=1$ model
\eqa
&& \quad S \, {\rm (nonl)} = {-i \over 2} \int dt d \theta G(\phi)
D_\theta \phi \del_t \phi \nn
&&= \int dt \left[ G (\varphi) \left( {1 \over 2} \dot\varphi^2 - {i
\over 2} \l_1 \dot\l_1 \right) + {i \over 2} \l_1 G' (\varphi) i \l_1
\dot\varphi \right] \nn
&& \quad = \int dt G (\varphi) \left( {1 \over 2} \dot\varphi^2  - {i
\over 2} \l_1 \dot\l_1 \right).
\eqae
Similarly, one can find nonlinear $\s$ models with $N=2$ susy in
$N=1$ superspace.  For all these $N=2$ models in $N=1$ superspace
there is also a $N=2$ superspace formulation.

The action in Hamiltonian form for the $N=2$ model reads
\eqa
&& L = \dot\varphi p + \dot\l^1 \pi^1 + \dot\l^2 \pi^2 + \dot{F} p_F
- {1 \over 2} F^2 - {1 \over 2} p^2,\nn
&& \delta \varphi = {i \over 2} \epsilon^j \left( \l^j + {2 \over i}
\pi^j \right), \quad \delta p =0,\quad \delta \pi^j = {i \over 2} p
\epsilon^j ,\nn
&& \delta \l^1 = p^1 \epsilon^1 + F \epsilon^2,\quad \delta \l^2 = p^2
\epsilon^2 - F \epsilon^1, \nn
&& \delta p_F = \epsilon^1 \pi^2 - \epsilon^2 \pi^1, \quad \delta F =0.
\eqae
As in the $N=1$ case, there are primary constraints $\pi^j - {i \over
2} \l^j =0$ and $p_F =0$, and secondary constraints
$F=0$.  Since $\delta F \sim \dot\l$ in the Lagrangian formulation,
but $\dot\l =0$ is the full field equation for $\l$, we
cannot express $\dot\l$ in terms of $\del_x \l$ by means of its field
equation, and this explains why $\delta F =0$ and
$\delta p =0$.  The susy
Noether charges in the Hamiltonian approach are given by
\eqa
Q^1_{H} &=& \left( i \pi^1 - {1 \over 2} \l^1 \right) p- iF \pi^2, \nn
Q^2_{H} &=& \left( i \pi^2 - {1 \over 2} \l^2 \right) p + i F \pi^1.
\eqae
These changes reproduce the Hamiltonian susy laws \emph{ exactly} if
one uses ordinary Poisson brackets.  The reason is that in
the Hamiltonian action there are no constraints.  In the action of
the Lagrangian formulation one finds Dirac brackets and the
following hermitian susy charges
\eqa
Q_L^j = - \l^j p.
\eqae
The algebra reads in both cases
\eqa
\{ Q^i_H , Q^j_H \}_P &=& \delta^{ij} (-p^2) =- 2 \delta^{ij} H, \nn
\{ Q^i_L, Q^j_L \}_D &=& \delta^{ij} (-p^2) =- 2 \delta^{ij} H.
\eqae
Thus $F$ and $p_F$ do not transform under $Q_L$, but $p_F$ transforms
under $Q_H$.  However, the algebra on all fields is the
same.  For example, $\epsilon Q_H \l \sim p \epsilon + F \epsilon
=0$ and $Q_L p = \epsilon$, in agreement
with $H \l =0$ for
$H= H_L = H_H$.

The $N=2$ model can be used as a toy model for instanton physics. In
Minkowski time $t$ the action is
\eqa
&& L = {1 \over 2} \dot\varphi^2 - i \bar\l \dot\l - \bar\l \l (m - g
\varphi) + {1 \over 2} \left( F - m \varphi + {1 \over
2} g \varphi^2 \right)^2 \nn &&\quad\quad - {1 \over 8} g^2 \left[ \left( \varphi - {m
\over g} \right)^2 - \left( {m \over g} \right)^2
\right]^2.
\eqae
This action is hermitian, and $\bar\l = (\l)^\dagger$.  In Euclidean
time $\tau, \l$ and $\bar\l$ become independent complex
spinors.  The Wick rotation is a complex Lorentz rotation (a $(U(1)$
rotation) in the $(t, \tau)$ plane (see the joint paper with Waldron)
\eqa
t^1_\theta = e^{i \theta} t, \;\; t^1_{\theta = \pi/2} \equiv \tau = i t.
\eqae
The spinor $\l$ transforms then half as fast
\eqa
\l^1_\theta = e^{i \theta/2} \l, \l_{\theta = \pi/2} \equiv \l_E = \sqrt{i} \l.
\eqae
Making these substitutions, one automatically obtains a
supersymmetric action for the Euclidean case.

It is a pleasure to thank Martin Ro\v{c}ek for discussions about the
covariant approach to superspace supergravity and the field redefinition in (\ref{2one17}).

\section{BRST quantization in a Hamiltonian approach}
So far we have been discussing classical actions. The classical
supergravity action was a gauge action, an action with two local
symmetries in out toy model: diffeomorphisms and local supersymmetry. In the
quantum theory, one uses a quantum action which is obtained by
adding two more terms: a gauge fixing term and a ghost action.
The local symmetry is then broken, but a rigid residual symmetry
remains, the so-called BRST symmetry (due to Becchi, Rouet, Stora
and Tyutin). The crucial property of BRST transformations is that
they are nilpotent (see below). An infinitesimal BRST
transformation has as parameter a purely imaginary anticommuting
constant $\L$. It is not the supersymmetry parameter (which is
also anticommuting). Physicists use (at least ) two ways to
formulate the BRST formalism: a Lagrangian approach and a
Hamiltonian approach. The infinitesimal BRST transformation rules
in the former are written as $\d_B\Phi$ (where $\Phi$ denotes any
field), but in the Hamiltonian approach one uses operators and
brackets. For example, the BRST transformations are generated by
a BRST operatorial charge $Q$, and for every field there is a
canonically conjugate field (called momentum by physicists) which
satisfy equal-time canonical commutation  rules or
anticommutation rules. Then one has $\d_B\Phi\sim\{\Phi,Q\}$.
Nilpotency means $\d_B\d_B\Phi=0$ in the Lagrangian approach, and
$\{Q,Q\}=0$ in the Hamiltonian approach. In this section we apply
this general formalism to our quantum mechanical toy model. One
word about terminology: we use the words real and hermitian (and
purely imaginary and antihermitian) as equivalent.

To quantize the supergravity action using covariant quantization
in the Lagrangian approach to BRST symmetry, we should add a gauge
fixing term and a corresponding ghost action. We begin with the
classical action in~(\ref{1thirtythree}) although we could also
start with the classical action in~(\ref{1sixtyfoour}). We fix the
gauge of general coordinate transformations by $h=0$, and the
gauge of local supersymmetry by $\psi =0$. The corresponding
gauge fixing terms in the action are then given by
\begin{equation}\label{pod} L_{\rm fix} = dh+\Delta\psi.
\end{equation}
The fields $d$ and $\Delta$ are Lagrange multipliers which fix
the gauges according to $h=0$ and $\psi=0$. Hermiticity of the
action requires that $d$ be real and $\Delta$ antihermitian. The
ghost action is then given by \eqa L {\rm (ghost)} \L = b \d_B h
+ \beta \d_B \psi, \eqae where $\d_B$ are the BRST transformations
and $\L$ is the constant anticommuting imaginary BRST parameter.
The fields $b$ and $\b$ are the antighosts; $b$ and $\b$ are
antihermitian. (Mathematical physicists remove $\L$ on both sides
of this equation and call it then a derivation. It is usually
denoted by $s$, and the relation to $\delta_B$ is
$\delta_B\phi=\Lambda s\phi$ for any field $\phi$). We obtain
thus for the full quantum action \eqa\label{brsttran} && L=L {\rm
(class)} + L {\rm (fix)} + L {\rm ghost)}, \nonumber \\ && L {\rm
(class)} = {1 \over 2} \dot\varphi^2 + {i \over 2} \l \dot\l -h
\dot\varphi^2 -i \psi \dot\varphi \l, \nonumber\\ && L {\rm (fix)}
= dh + \Delta \psi,\quad L  {\rm (ghost)} \L = b \d_B h + \beta \d_B
\psi. \eqae

For the classical fields ($h$ and $\psi$ and $\varphi$ and $\l$)
the BRST transformations are just gauge transformations with a
special choice of the parameters: $\xi = c\L$ and $\epsilon =- i
\g \L$ where $c$ and $\g$ are real ghosts ($\L$ is imaginary, and
$\xi$ and $\epsilon$ are real). Because $\xi$ is commuting, $c$ is anticommuting, 
and because $\epsilon$ is anticommuting, $\gamma$ is commuting. 
The BRST transformation rules for
the classical fields follow from~(\ref{1thirtythree}),
(\ref{varphi}) and~(\ref{dot}) \eqa && \d_B h = [(1-2 h) \psi \g
+ {1 \over 2} (1-2 h) \dot{c} + \dot{h} c ] \L, \nn && \d_B \psi =
[ -i (1-2 h) \dot\g + \dot\psi c ] \L, \nn && \d_B \varphi = (- \l
\g + \dot\varphi c) \L,\quad \d_B \l = [ i (1-2 h) \dot\varphi \g+
\psi\l\g + \dot\l c ] \L .\label{class}\eqae
The ghost action then becomes
\begin{gather}
L(\mathrm{ghost})=b[(1-2h)\psi\gamma +\frac{1}{2}(1-2h)\dot{c} +\dot{h}c] 
+ \beta[-i(1-2h)\dot{\gamma}+ \dot{\psi}c].
\end{gather}
It is clearly hermitian.

The BRST transformation rules of the ghosts follow from the
structure constants of the classical
 gauge algebra or from the nilpotency of BRST transformations.  One has uniformly on all classical
fields ($\varphi , \l, h$ and $\psi$), as discussed
in (3.18)
\eqa && [ \d_s (\e_2) ,
\d_s (\e_1) ] = \d_g (2i (1-2 h) \e_2 \e_1 ) + \d_s (-2 i \psi
\e_2 \e_1 ), \nn && [ \d_g ( \xi_2 ), \d_g (\xi_1) ] = \d_g (-
\xi_2 \dot{\xi_1} + \xi_1 \dot\xi_2 ), \nn && [ \d_g (\xi) , \d_s
(\e) ] = \d_s  (- \xi \dot\e ). \eqae Thus the classical local
gauge algebra ``closes":  the commutator of two local symmetries
is a linear combination of local symmetries.  New from a
mathematical point of view is the appearance of fields ($h$ and
$\psi$) in the composite parameters, and thus in the structure
constants.  (One should thus rather speak of ``structure
functions").  One obtains \eqa \d_B c &=& -c \dot{c} \L +i (1-2
h) \g \g \L ,\nn \d_B \g &=& c \dot\g \L +  \psi \g \g \L .\eqae An
easier way to obtain these results is to use that the BRST
variation of~(\ref{class}) should vanish. One may check that all
BRST transformation rules preserve the reality properties of the
fields.

The antighosts $b$ and $\beta$ transform into the auxiliary
fields $d$ and $\Delta$, and the auxiliary fields are BRST
invariant (``contractible pairs") \eqa\label{contr} \d_B b= \L d
, \quad  \d_B \beta =\L \Delta , \quad \d_B d=0 ,\quad  \d_B \Delta =0. \eqae All
BRST transformation laws are nilpotent, and they leave the action
$S=\int L dt$ invariant.

In the Lagrangian approach the BRST charge $Q$ is the Noether
charge for rigid BRST transformations but one does not use
brackets. To obtain $Q$, one lets $\L$ become local (t-dependent),
and one collects all terms in the variation of the action
proportional to $\dot\L$. (We used this procedure before to
construct the supersymmetry charge). The BRST transformation
rules of the fields themselves for local $\L$ do not contain by
definition any $\dot\L$; for example, $\d_B \psi = (1-2 h) (-i
\dot\g) \L (t)+\dots$, and not $\d_B \psi = (1-2 h) {d \over dt}
(-i \g \L (t))+\dots$. The classical action yields the following
terms proportional to $\dot\L$ \eqa \d S {\rm (class)} = \int [ L
{\rm (class)}  c \dot\L - (1-2 h) \l \dot\varphi \g \dot\L ] dt.
\eqae The term with $c \dot\L$ is expected from the result $\d
\cl = \del_\a ( \xi^\a \cl )$ in general relativity, while the
term with $\g \dot\L$ comes from the variation with ${d \over dt}
\e$ in $i \l {d \over dt} \d \l$ which is canceled by the Noether
term.  The gauge fixing term produces no terms with $\dot\L$
because $\d_B h$ and $\d_B \psi$ do not contain $\dot\L$ terms by
definition, but the ghost action
 yields further terms proportional to $\dot\L$
\eqa && \d S {\rm (ghost)} =  b [ {1 \over 2} (1-2 h) ( -c
\dot{c} +i (1-2 h) \g \g ) \dot\L \nn &&  \qquad + c ( (1-2h )
\psi \g + {1 \over 2} (1-2 h) \dot{c} + \dot{h} c ) \dot\L ] \nn
&& + \beta [  (-i) (1-2h) (c \dot\g +  \psi \g \g ) \dot\L + i c
(1-2h) \dot\g \dot\L ] \eqae Several terms cancel in this
expression.

The BRST charge in the Lagrangian approach is thus \eqa Q &=& c L
{\rm (class)} - (1-2 h) \l \dot\varphi \g \nn &+& b [  {i \over
2} (1-2 h)^2 \g \g + (1-2 h) c \psi \g ] \nn &-& i\beta (1-2 h)
\psi \g \g. \label{is} \eqae

In the Hamiltonian approach the BRST charge should contain terms
of the form (we discuss this in more detail later) \eqa Q = c^\a
\phi_\a-\frac12c^\b c^\a f_{\a\b}{}^\g b_\g(-)^\b, \eqae with
ghosts $c^\a = (c, \g)$, antighosts $b_\a = (b, \beta)$, and
first class constraints $\phi_\a = (T, J)$ where $T$ is the
generator of diffeomorphisms and $J$ the generator of
supersymmetry. The sign $(-)^\beta$ is equal to $+1$ if the
corresponding symmetry has a commuting parameter (and thus an
anticommuting ghost); when the ghost is commuting $(-)^\beta$
equals $-1$. However, because the structure constants contain
$\dot\xi$ and $\dot \epsilon$ we expect in $Q$ terms with
derivatives of $c$, namely $bc \dot{c}$ and $\b \dot\g c$ terms.
On the other hand, in a truly Hamiltonian approach no time
derivatives of fields are allowed in the charges. This suggests that the $b$
and $\beta$ field equations have been used to eliminate $\dot{c}$
\eqa \dot{c} = - 2 \psi \g - {2 \over 1-2 h} \dot{h} c;\qquad
\dot\g = -\frac{i\dot\psi c}{1-2h}. \eqae However, then one
obtains time derivatives of $h$ and $\psi$. Another problem is
that we seem to have too many factors of $(1-2 h)$ but this could
be repaired by redefining fields. To resolve these issues we
first construct $Q$ by Hamiltonian methods. This is a very
general approach which only uses as input the first-class
constraints of the classical theory, and which provides a quantum
action in phase space and a Hamiltonian BRST charge with in
general many more fields than in the usual (Lagrangian)
formulation. Eliminating nonpropagating fields we should regain
the Lagrangian BRST charge $Q$ in~(\ref{is}). Let's see how this
works out.

In the Hamiltonian framework of Fradkin and Vilkovisky (and
others) the quantum action is of the form \eqa L = \dot{q}^i p_i
- H + \{ Q_H, \psi_g \}, \label{Ham} \eqae where $q^i$ denotes all
fields: classical fields (including the Lagrange multipliers $h$
and $\psi$), ghosts and antighosts.  The $p_i$ are canonical
momenta \emph{for all of them}. (So, for example, there are
canonical momenta for the ghosts, and separate canonical momenta
for the antighosts). The BRST charge $Q_H$ should be nilpotent
\eqa \{ Q_H , Q_H \} =0. \eqae The full quantum Hamiltonian $H$ is
constructed from Dirac's Hamiltonian $H_D$ such that
\begin{equation}
[H, Q_H] =0.
\end{equation}
Neither $H_D$ nor $H$ should depend on Lagrange multipliers, and
will be constructed below. The BRST charge is in general given by
\eqa\label{ingen} Q_H = c^\a \varphi_\a +  p^\mu (B) \pi_\mu (\l)
- {1 \over 2} c^\beta c^\a f_{\a\b}{}^\g p_\g (-)^\beta ,\eqae
where $\varphi_\a$ are the first-class constraints,
$f_{\a\b}{}^\g$ the structure functions defined by
$\{\varphi_\a,\varphi_\b\}=f_{\a\b}{}^\g\varphi_\g$, and $\l^\mu$
are the Lagrange multipliers (classical fields which appears in
the classical action without time derivatives), and $B$ denotes
the antighosts. The structure functions should only depend on
$p_i$ and $q^i$ but not on $\l^\mu$. Usually one has to take
suitable linear combinations of local symmetries and add
so-called equation-of-motion symmetries to achieve this. We shall
demonstrate this in our model.

The BRST invariance of the action in~(\ref{Ham}) is almost
obvious: each of the 3 terms is separately invariant due to the
relations $\int \frac{d}{dt}Q_H dt =0$, $[Q_H,H]=0$ and
$\{Q_H,Q_H\}=0$.

We start again from the classical gauge invariant action
in~(\ref{1thirtythree})
\begin{equation}
L=\frac12 (1-2h){\dot\varphi}^2+\frac{i}{2}\lambda {\dot \lambda} -
i\psi{\dot \varphi}\lambda.
\end{equation}
The primary constraints are $p_h=0$, $\pi_{\psi}=0$ and
$\pi_{\lambda}+ \frac{i}{2}\lambda=0$, and the naive Hamiltonian
with all primary constraints added, reads
\begin{equation}
H_{\rm naive}=\frac{1}{1-2h}\frac12\left( p+i\psi\lambda\right)^2+
ap_h+\alpha\pi_{\psi}+\eta(\pi_{\lambda}+\frac{i}{2}\lambda).
\end{equation}
The functions $a(t)$, $\alpha(t)$ and $\eta (t)$ are arbitrary
Lagrange multipliers which enforce the primary constraints. By a
redefinition of $\eta$ we can replace $\lambda$ in the first term
by the field $\frac12 \lambda+i\pi_{\lambda}$ which anticommutes
with the constraint $\pi_{\lambda} +\frac{i}{2}\lambda $. This
will simplify the analysis.

Conservation of the 3 primary constraints yields two secondary
constraints and fixes one Lagrange multiplier
\begin{equation}
p^2=0,\quad p(\pi_{\lambda} -\frac{i}{2}\lambda)=0,
\quad \eta (t) =0.
\end{equation}
Both secondary constraints are preserved in time
because $\pi_{\lambda} -\frac{i}{2}\lambda$ anticommutes with the constraint
$\pi_{\lambda} +\frac{i}{2}\lambda$. The Hamiltonian can be rewritten as
\begin{equation}
H_{\rm naive}=(1+2H)\frac12 p^2+i\Psi(\frac{\lambda}{2} + i\pi_{\lambda})
p+ap_h+\alpha\pi_{\psi}.
\end{equation}
Here
\begin{equation}\frac{1}{1-2h}=1+2H \quad\text{and}\quad
\frac{\psi}{1-2h}=\Psi,
\end{equation}
so we encounter again the fields $H$ and $\Psi$
of~(\ref{dvatr}). From now on we use $G\equiv1+2H$ as
gravitational field, and $\Psi$ as gravitino. Hence, the Dirac
Hamiltonian  (the Hamiltonian on the constraint surface) vanishes
\begin{equation}
H_{D}=0.
\end{equation}
and there are 5 constraints: one second class constraint
\begin{equation}\label{secclass}
\pi_{\lambda} +\frac{i}{2}\lambda=0,
\end{equation}
and four first-class constraints
\begin{equation}\label{four1}
p_h=\pi_{\psi}=p^2=p(\pi_{\lambda} -\frac{i}{2}\lambda)=0.
\end{equation}
Thus the Dirac brackets in (2.19) remain valid for supergravity.

The constraint $p^2=0$ generates diffeomorphisms. One might expect
that they are given by \eqa \delta\varphi=\xi{\dot \varphi}, \quad
\delta p=\xi{\dot p}, \quad \delta\lambda=\xi{\dot \lambda},\quad
\delta\pi=\xi{\dot \pi}, \nonumber \\
\delta H= \frac{d}{dt}\left[ \frac12(1+2H)\xi\right],\quad
\delta\Psi = \frac{d}{dt}(\xi\Psi),\label{delhal} \eqae because
these transformations leave~(\ref{1Gaussian}) invariant. However
in the Hamiltonian tranformation laws, no time derivatives are
allowed. Moreover one expects that $p^2$ can only act on
$\varphi$ (and perhaps $p$ and $H$), but not an $\l$, $\pi$ and
$\psi$. We now perform a series of modifications of the
transformation rules which cast~(\ref{delhal}) into the expected
form.

By adding terms proportional to the $p$ and $\varphi$ field
equations to $\d\varphi$ and $\d p$, respectively (so-called
equation of motion symmetries), namely, $\delta\phi=-\xi\tfrac{\del S}{\del p}$ 
and $\delta p=\xi\tfrac{\del S}{\del \phi}$, one obtains \eqa
&&\delta\varphi  = (1+2H)\xi p - \xi \Psi (\pi - \frac{i}{2}\lambda),\nonumber \\
&&\delta p  =0. \eqae Similarly, adding terms proportional to the $\pi$ and $\l$ field
equations to $\delta \lambda$ and $\delta \pi$ yields
\begin{equation}
\delta\lambda = - \xi p \Psi, \quad \delta\pi_\l = \frac{i}{2}\xi
p \Psi.
\end{equation}
As a check of these last two results note that the constraint
$\pi_\l + \frac{i}{2}\lambda$ is invariant. Finally we can remove
the complicated last term in $\delta\varphi$ adding a local susy
transformation in~(\ref{1Gaussian}) with parameter $\epsilon =-
\xi\Psi$ to {\bf all} fields. This yields \eqa\label{pdv}
\delta\varphi={\hat\xi}p, \quad \delta p=0, \quad
\delta\lambda=0,\quad
\delta\pi_\l=0, \nonumber\\
\delta\Psi=0,\quad \delta (1+2H)= \frac{d}{dt}{\hat \xi},\quad
{\hat \xi} = (1+2H)\xi. \eqae These are the transformations of the
matter fields generated by $\frac{1}{2}p^2$. The classical gauge
fields are $(1+2H)$ and $\Psi$ and they transform in general as
\begin{equation}\label{AAA}
 \delta h^A = \frac{d}{dt} \epsilon^A + f^A_{\ BC} h_{\mu}^B \epsilon^C,
\end{equation}
where $\epsilon^A$ and $h^A_\mu$ correspond to $\hat \xi$ and
$1+2H$ in the case of diffeomorphisms. The same results should be
obtained by taking the brackets with $Q_H$.

Next we consider the local susy generator in~(6.20) which
we already multiply with the classical susy gauge parameter
$\epsilon (t)$, hence $\epsilon (\pi_{\lambda}
-\frac{i}{2}\lambda)p$. It generates the following classical transformation
laws, obtained using the Dirac brackets,
\begin{equation}\label{pdv2}
\delta\varphi =- \epsilon (\pi_{\lambda} -\frac{i}{2}\lambda),\ \ \
\delta p = 0, \ \ \
\delta\lambda=-p\epsilon,\ \ \
\delta\pi_{\lambda}=\frac{i}{2}p\epsilon.
\end{equation}
These are the transformation laws of~(\ref{1Gaussian}).
The gauge fields should transform according to~(\ref{AAA})
\begin{equation}\label{pdv3}
\delta\Psi = {\dot \epsilon },\qquad \delta (1+2H)=
-2i\epsilon\Psi.
\end{equation}
(The factor 2 in $-2i\epsilon\Psi$ comes from the two terms in
$f^{A}_{\ \ BC}h_{\mu}^B\epsilon^C$ with $h_{\mu}^B=\Psi$ and
$\epsilon^C=\epsilon$, or vice-versa). Also these rules agree
with~(\ref{1Gaussian}). (Note that the local classical gauge
algebra based on~(\ref{pdv}), (\ref{pdv2}) and (\ref{pdv3})
closes, and that only the commutator of two local supersymmetry
transformations is nonzero).

The local gauge algebra of the transformation in~(\ref{pdv})
and~(\ref{pdv2}),~(\ref{pdv3}) has now structure functions which
are independent of the Lagrange multipliers $h$ and $\psi$ (or
$H$ and $\Psi$), just as required for a Hamiltonian treatment.
Having shown that the two first class constraints $\tfrac{1}{2}p^2$ and
$p(\pi_{\lambda} -\frac{i}{2}\lambda)$ generate indeed the local
symmetries of the classical phase space action
in~(\ref{1Gaussian}), we now proceed with the construction of the
BRST charge $Q_H$ and the quantum action.

The BRST charge is given by \eqa\label{pdv4} Q_H=\frac12c p^2 -
i\gamma p(\pi_{\lambda} -\frac{i}{2}\lambda)
+p_bp_G+\pi_{\beta}\pi_{\Psi} - i\pi_c\gamma\gamma, \eqae where
$G=1+2H$. We denote antihermitian conjugate momenta by $\pi$ as
in $\pi_{\lambda}$, $\pi_{\Psi}$, $\pi_c$, $\pi_{\beta}$, but
hermitian momenta by $p$ as in $p_G$, $p_b$, $p_\g$. By $p$
in~(\ref{pdv4}) we mean $p_{\varphi}$, as before. (Recall that all
ghosts are real, but the antighosts $b$ and $\b$ are
antihermitian). The BRST charge is real and anticommuting. The
first four terms contain the four first-class constraints, and
the last term in~(\ref{pdv4}) comes from the last term
in~(\ref{ingen}). The coefficients of the terms with $p_b$ and
$\pi_\b$ are not fixed by nilpotency. It is easiest to fix the
coefficients and signs by working out the transformation rules
and fitting to the results obtained earlier, although in
principle we need not follow this path since all terms are well
defined. One can prove the nilpotency of the BRST laws by
directly evaluating $\{Q_H,Q_H\}$, using the equal-time canonical
(anti)commutations relations.

Defining BRST transformations by
\begin{equation}\label{ptr}
\delta_{B}\Phi = -i[\Lambda Q_H,\Phi]=-i[\Phi,Q_H\L],
\end{equation}
 for any field $\Phi$, we find the following results
\begin{equation}\label{delb}
\begin{array}{l}
\delta_B\varphi = c p \Lambda - i (\pi_{\lambda} -\frac{i}{2}\lambda)\gamma\Lambda, \\
\delta_{B}\lambda = i\gamma p \Lambda, \\
\delta_{B}G = p_b \Lambda,\\
\delta_{B} c = i\gamma \gamma \Lambda,\\
\delta_{B}\gamma=0,\\
\d_B\pi_c =-\frac12 p^2 \L\\
\d_B p_G = \d_B\pi_\psi = \d_Bp_b=\d_B\pi_\b = 0,
\end{array}
\begin{array}{l}
\delta_{B}p=0,\\
 \delta_{B}\pi_{\lambda} =\frac12 \g p \L, \\
\delta_B{\Psi}=-\pi_{\beta} \Lambda, \\
\delta_{B}b=-{\Lambda} p_G, \\
\delta_{B}\beta=\pi_\psi\Lambda, \\
\d_B p_\g
=2i\pi_c\g\L+ip(\pi_\l-\frac{i}{2}\l).
\end{array}
\end{equation}
On the classical fields these rules agree with the classical gauge
transformations with $\xi=c\L$ and $\epsilon=-i\g\L$. In
principle one should use Dirac brackets to obtain these
transformation rules. These Dirac brackets can be constructed
from the second class constraint in~(\ref{secclass}), but because
the second class constraint commutes with the first class
constraints, the results are the same\footnote{
Only in the exceptional case that the Poisson bracket of a first
and a second class constraint gives a square of a second class
constraint, one would get a different answer, and in that case
one would need to use Dirac brackets. This situation does not
occur in our model.}.

The rules in~(\ref{delb}) are nilpotent. In fact, we could have
used nilpotency of $Q_H$ to derive $\delta_B c$ and
$\delta_B\gamma$ from $\delta_B\varphi$, namely as follows \eqa
&&\delta_B^2\varphi =0 = \delta_B\left[ cp-i\gamma(\pi_{\lambda} -
\frac{i}{2}\lambda)\right]\nonumber\\
&&\qquad =(\delta_B c)p - i(\delta_B\gamma)(\pi_{\lambda}
-\frac{i}{2}\lambda)- i\gamma (\gamma
p\Lambda)\;\Rightarrow\;\delta_B c =i\gamma\gamma\Lambda,\ \
\delta_B\gamma=0. \eqae The result $\delta_B\gamma=0$ can also be
derived from $\delta_B^2\lambda=0$.

Comparison of $\d_Bb$ and $\d_B\b$ in~(\ref{delb})
and~(\ref{contr}) reveals that the BRST auxiliary fields are the
canonical momenta of the Lagrange multipliers
\begin{equation}
p_G=-d,\qquad \pi_{\Psi}=-\Delta.
\end{equation}

Finally we construct the quantum action in the Hamiltonian approach
\begin{equation}
L={\dot q}^ip_i-H+\{ Q_H,\psi_g\}.
\end{equation}
As ``gauge-fixing fermion'' we take the following hermitian anticommuting expression
\begin{equation}\label{gaugefix}
\psi_g = -i(b(G-1)+G\pi_c)+(-i\beta\Psi-\Psi p_{\gamma}).
\end{equation}

The quantum Hamiltonian $H_H$ which commutes with $Q_H$ vanishes in our case
(and in any gravitational theory) because $H_D=0$ (see (6.23)). We denote the gravitational field
$1+2H$ by $G$, and find then for the quantum Hamiltonian
\begin{equation}\label{acti}
L={\dot \varphi}p +{\dot \lambda}\pi_{\lambda}+{\dot G}p_G + {\dot
\Psi}\pi_{\Psi} +{\dot c}\pi_c +{\dot b}p_b+{\dot
\gamma}p_{\gamma} + {\dot \beta}\pi_{\beta}+\{Q_H, \psi_g\}.
\end{equation}
The evaluation of $\{Q_H,\psi_g\}$ is tedious but
straightforward. For any pair of canonically conjugate variables
we have $[p,q]=\frac1i$ or $\{\pi ,q\}=\frac1i$. This yields
\eqa\label{esse} \{Q_H,-
i(b+\pi_c)G+(-i\beta-p_{\gamma})\Psi\}&&\!\!\!\!\!\!\!\!=
-p_G(G-1)-\frac12 Gp^2+(b+\pi_c)p_b-\pi_\Psi\Psi
 \nonumber\\
&&\!\!\!\!\!\!\!\!- p(\pi_{\lambda} -\frac{i}{2}\lambda)\Psi -
2\pi_c\gamma\Psi+\pi_\b(-\b+ip_\g). \eqae The transformation laws
$\d_BG=p_b\L$ and $\d_B\Psi=-\pi_\b\L$ which follow
from~(\ref{ptr}) agree with the rules $\d_BG=({\dot
c}+2\g\Psi)\L$ and $\d\Psi=-i{\dot \gamma}\L$ which we found from
the classical transformation laws by substituting $\xi=c\L$ and
$\epsilon=-i\g\L$, provided
\begin{equation}\label{pigam}
-p_b+{\dot c}+2\g\Psi=0, \qquad i\pi_\b+{\dot \g} =0.
\end{equation}
These should be algebraic field equations, and indeed they are
the field equations of $\pi_c$ and $p_\g$. The relevant terms in
the action are
\begin{equation}
L=(-p_b+{\dot c}+2\g\Psi)\pi_c + ({\dot \g}+i\pi_\b)p_\g.
\end{equation}
Integrating out $\pi_c$, $p_b$, $p_\g$ and $\pi_\b$
imposes~(\ref{pigam}).

The terms $-p_G(G-1)$ and $-\pi_{\Psi}\Psi$ are the gauge fixing
terms. Thus to compare with the action as obtained from the
Lagrangian BRST formalism we should use $b(G-1)+\Delta\Psi$
in~(\ref{brsttran}) as gauge fixing term . However, we now got
terms ${\dot G}p_G-p_GG$ and ${\dot \Psi}\pi_\Psi-\pi_\Psi\Psi$
in the action. In Yang-Mills gauge theories one usually takes
$\psi_g=b_a\chi^a+p_{c,a}A_0^a$ where $A_0^a$ is the
time-component of the classical gauge field, and
$\chi^a=\partial^kA_k^a$. Then one makes a change of integration
variables $p(A_0)=kp(A_0)'$ and $b_a=kb_a'$ where $k$ is a
constant. The superjacobian is unity, and taking the limit
$k\rightarrow 0$ one arrives at the quantum action with
relativistic unweighted gauge
$\partial^{\mu}A_{\mu}^a=0$~\cite{Henne}. In our case the gauge
choices were $G-1=0$ and $\Psi=0$. We used the same $\psi_g$ but
with both $\chi^a$ and $A^a_0$ equal to $2H$ and $\Psi$. This led
to~(\ref{gaugefix}), but it is clear that we cannot rescale
$A_0^a\sim(G$ and $\Psi)$ but keep fixed $\chi\sim(G$ and
$\Psi)$. In fact, we could have noticed before that there is
something wrong with $\psi_g$ in~(\ref{gaugefix}): the terms have
different dimensions.

The resolution is also clear: drop the terms with $b$ and $\b$
in~(\ref{gaugefix}). This removes the offending terms
$-p_GG-\pi_\Psi\Psi$ and also the terms $bp_b$ and
$-\pi_\beta\beta$ in~(\ref{esse}). The action now becomes after
eliminating $\pi_c$, $p_b$, $\pi_\g$ and $\pi_b$ \eqa
\label{theac} L&\!\!=&\!\!{\dot \varphi} p + {\dot \l} \pi_\l
-\frac12
Gp^2-p(\pi_\l-\frac{i}{2} \l)\Psi\nonumber \\
&+&\!\! {\dot G}p_G+{\dot \Psi}\pi_\Psi\\
&+&\!\! {\dot b} ({\dot c}+2\g\Psi)+{\dot\b}(i{\dot \g}).
\nonumber\eqae The first line is~(\ref{1Gaussian}), the second
line is the gauge fixing term, and the third line the ghost
action. So the action in~(\ref{acti}) from the Hamiltonian BRST
formalism indeed agrees with the action from the Lagrangian BRST
formalism if one chooses ${\dot G}$ and ${\dot \Psi}$ instead of
$G$ and $\Psi$ as gauge fixing terms.

We have seen that there is a Lagrangian and a Hamiltonian
approach. The latter contains conjugate momenta for all
variables, hence brackets can be defined and charges constructed.
We saw that the gauges $G-1=0$ and $\Psi=0$ in the Hamiltonian
approach led to the same results after solving the algebraic
field equations for the canonical momenta as the gauge ${\dot
G}=0$ and ${\dot  \Psi}=0$ in the Lagrangian approach. (We recall
that $G=1+2H=\frac{1}{1-2h}$ and $\Psi=\frac{\psi}{1-2h}$.) One
might ask whether the Hamiltonian approach can also lead to the
gauge $G-1=0$ and $\Psi=0$ in the Lagrangian approach
(corresponding to $h=0$ and $\psi =0$ in~(\ref{pod})). This is
indeed possible. One chooses as gauge fermion $\psi_g=0$. The
action becomes then very simple \begin{equation}\label{simpac}
L=\dot{\varphi}p+\dot{\l}\pi_\l+\dot{G}p_G+\dot{\Psi}p_{\Psi}+
\dot{c}\pi_{c}+\dot{b}p_b+\dot{\g}p_\g + \dot{\b}\pi_\b.
\end{equation}
Next one factorizes the path integral into a minimal part with the
fields $(p,\varphi)$, $(\l,\pi)$, $(c,\pi_c)$ and $(\g,p_\g)$,
and a nonminimal part with the pairs $(G,p_G)$, $(\Psi,\pi_\Psi)$,
$(b,p_b)$ and $(\b,\pi_\b)$. Finally one discards the latter, and
\emph{reinterprets} the momenta $\pi_c$ and $p_\g$ as the
Lagrangian antighosts $b$ and $\b$, respectively. This yields
then the action in~(\ref{brsttran}) with $d$, $\Delta$, $h$ and
$\psi$ integrated out (removed by their algebraic field
equations).

This concludes our discussion of BRST formalism applied to our
simple model. We obtained one result which is a bit surprising
(or interesting): the differentiated gauge choices
$\dot{G}=\dot{\Psi}=0$ in the Lagrangian approach correspond
directly to the gauge choices $G-1=\Psi=0$ in the Hamiltonian
approach. On the other hand, the gauge choices $G-1=\Psi=0$ in
the Lagrangian approach did not correspond in a direct way to the
Hamiltonian approach (we had to discard a sector). String theory
uses Lagrangian gauge choices corresponding to $G-1=\Psi=0$.
Perhaps the corresponding differentiated gauge choices have
advantages in certain respects.

\end{document}